\documentclass[11pt]{article}
\usepackage{amsmath}
\usepackage{hyperref}
\usepackage{wasysym}
\usepackage[font=footnotesize]{caption}
\usepackage[T1]{fontenc}
\usepackage{latexsym}
\usepackage[english]{babel}
\usepackage{cite}
\usepackage{enumerate}
\usepackage[shortlabels]{enumitem}
\pdfoutput=1
\usepackage[utf8]{inputenc}
\usepackage{dsfont}
\usepackage{diagbox}
\usepackage{amsfonts}
\usepackage{mathtools}
\allowdisplaybreaks[4]         
\usepackage{amssymb}
\usepackage{euscript}     
\usepackage{braket}
\usepackage{amssymb}
\usepackage{starfont}
\usepackage{color,soul}         
\usepackage{braket}
\usepackage{tensor}        
\usepackage{amsthm}
\usepackage{graphicx}
\usepackage{slashed}
\usepackage{leftidx}
\usepackage{subfigure}
\usepackage{bbm}
\usepackage{empheq}
\usepackage[dvipsnames]{xcolor}
\definecolor{outerspace}{rgb}{0.25, 0.29, 0.3}
\definecolor{scarlet}{rgb}{1.0, 0.13, 0.0}
\usepackage[header,title,page,titletoc]{appendix}  
\definecolor{princetonorange}{rgb}{1.0, 0.56, 0.0}
\definecolor{WildStrawberry}{rgb}{1.0, 0.26, 0.64}
\definecolor{rossocorsa}{rgb}{0.83, 0.0, 0.0}
\definecolor{navyblue}{rgb}{0.0, 0.0, 0.5}
\usepackage[numbers,sort&compress]{natbib}  
\usepackage{float}
\usepackage[paper=a4paper,margin=1in]{geometry}
\usepackage{titlesec}

\newcommand{\fo}{\tilde{F}}
\newcommand{\fii}{F_1}
\newcommand{\go}{\tilde{G}}
\newcommand{\gii}{G_1}

\pdfoutput=1
\usepackage{amsmath}
\usepackage{color}
\usepackage{amsfonts}
\usepackage{amssymb}
\usepackage{graphicx}
\usepackage{geometry}
\usepackage{amssymb,epsfig,subfigure}
\usepackage{amssymb}
\usepackage{hyperref}
\usepackage{comment}
\usepackage[font=footnotesize]{caption}
\usepackage[T1]{fontenc}
\usepackage[utf8]{inputenc}
\usepackage{latexsym}

\usepackage{cancel}

\usepackage[numbers,sort&compress]{natbib}  
\usepackage{float}

\usepackage{dsfont}
\usepackage{diagbox}

\usepackage{mathtools}
\allowdisplaybreaks[4]

\usepackage{euscript}     
\usepackage{braket}
\usepackage{starfont}
\usepackage{color,soul}         
\usepackage{tensor}        
\usepackage{amsthm}

\usepackage{slashed}
\usepackage{leftidx}
\usepackage{bbm}

\makeatletter
\renewcommand\section{\@startsection {section}{1}{\z@}%
                                 {-3.5ex \@plus -1ex \@minus -.2ex}
                                   {2.3ex \@plus.2ex}%
                                   {\normalfont\large\bfseries}}
\renewcommand\subsection{\@startsection{subsection}{2}{\z@}%
                                   {-3.25ex\@plus -1ex \@minus -.2ex}%
                                     {1.5ex \@plus .2ex}%
                                     {\normalfont\bfseries}}
\renewcommand\subsubsection{\@startsection{subsubsection}{3}{\z@}%
                                   {-3.25ex\@plus -1ex \@minus -.2ex}%
                                     {1.5ex \@plus .2ex}%
                                     {\normalfont\itshape}}
\makeatother

\def\pplogo{\vbox{\kern-\headheight\kern -29pt
\halign{##&##\hfil\cr&{\ppnumber}\cr\rule{0pt}{2.5ex}&\ppdate\cr}}}
\makeatletter
\def\ps@firstpage{\ps@empty \def\@oddhead{\hss\pplogo}%
  \let\@evenhead\@oddhead 
}
\def\maketitle{\par
 \begingroup
 \def\thefootnote{\fnsymbol{footnote}}
 \def\@makefnmark{\hbox{$^{\@thefnmark}$\hss}}
 \if@twocolumn
 \twocolumn[\@maketitle]
 \else \newpage
 \global\@topnum\z@ \@maketitle \fi\thispagestyle{firstpage}\@thanks
 \endgroup
 \setcounter{footnote}{0}
 \let\maketitle\relax
 \let\@maketitle\relax
 \gdef\@thanks{}\gdef\@author{}\gdef\@title{}\let\thanks\relax}
\makeatother

\numberwithin{equation}{section}

\newcommand\eea{\end{eqnarray}}
\newcommand\bea{\begin{eqnarray}}

\def\beq{\begin{equation}}
\def\eeq{\end{equation}}

\newcommand{\be}{\begin{equation}}
\newcommand{\ee}{\end{equation}}
\newcommand{\ba}{\begin{align}}
\newcommand{\ea}{\end{align}}
\newcommand{\bg}{\begin{gather}}
\newcommand{\eg}{\end{gather}}
\newcommand{\bseq}{\begin{subequations}}
\newcommand{\eseq}{\end{subequations}}

\usepackage{hyperref}
\hypersetup{
    colorlinks,
    citecolor=rossocorsa,
    filecolor=navyblue,
    linkcolor=navyblue,
    urlcolor=navyblue
}

\begin{document} 

\begin{titlepage}

\begin{center}

\phantom{ }
\vspace{3cm}

{\bf \Large{Charges in the UV completion of neutral electrodynamics}}
\vskip 0.5cm
Valentin Benedetti${}^{\dagger}$, Horacio Casini${}^{*}$, Javier M. Mag\'an${}^{\ddagger}$
\vskip 0.05in
\small{ ${}^{\dagger}$ ${}^{*}$ ${}^{\ddagger}$ \textit{Instituto Balseiro, Centro At\'omico Bariloche}}
\vskip -.4cm
\small{\textit{ 8400-S.C. de Bariloche, R\'io Negro, Argentina}}
\vskip -.10cm
\small{${}^{\ddagger}$ \textit{David Rittenhouse Laboratory, University of Pennsylvania}}
\vskip -.4cm
\small{\textit{ 209 S.33rd Street, Philadelphia, PA 19104, USA}}
\vskip -.10cm
\small{${}^{\ddagger}$ \textit{Theoretische Natuurkunde, Vrije Universiteit Brussel (VUB)}}
\vskip -.4cm
\small{\textit{ and The International Solvay Institutes}}
\vskip -.4cm
\small{\textit{ Pleinlaan 2, 1050 Brussels, Belgium}}

\begin{abstract}
A theory with a non-compact form-symmetry is described by two closed form fields of  degrees $k$ and $d-k$. Effective theory examples are non-linear electrodynamics, a photon field coupled to a neutron field, and a low energy Goldstone boson. We show these models cannot be completed in the UV without breaking the non-compact form-symmetry down to a compact one. This amounts to the existence of electric or magnetic charges. A theory with an unbroken non-compact $k$-form symmetry is massless and free.     
\end{abstract}
\end{center}

\small{\vspace{7.0 cm}\noindent ${}^{\dagger}$valentin.benedetti@ib.edu.ar\\
${}^{\text{\text{*}}}$casini@cab.cnea.gov.ar\\
${}^{\ddagger}$ magan@sas.upenn.edu
}

\end{titlepage}

\setcounter{tocdepth}{2}

{\parskip = .4\baselineskip \tableofcontents}
\newpage

\section{Introduction}
An electromagnetic field can have neutral interactions when the Lagrangian is written in terms of the field strength and it does not involve the vector potential. Examples are a higher derivative four photon term\footnote{Higher derivative theories of the field strength have been called ``Non-linear Electrodynamics'', see \cite{Russo:2022qvz, Guerrieri:2022sod}  and references therein for recent literature on the subject. An important example in this class of theories is Born-Infeld electrodynamics \cite{Born:1934gh}. Below we consider this class of theories together with magnetic couplings with neutral fields. We call this larger zoo of theories ``neutral electrodynamics''.} or the magnetic moment coupling to a neutral fermion:
\be
{\cal L} = -\frac{1}{4}\, F_{\mu\nu}  \, F^{\mu\nu}\, + \bar{\psi} (i \gamma^\mu \partial_\mu -m) \psi- 
 \frac{\mu_1}{8}\,(F^2)^2 -\,\frac{\mu_2}{2}\, \bar{\psi}\,\sigma_{\mu\nu}\,\psi\, F^{\mu\nu}\,. 
\label{lagra}
\ee
These couplings are non renormalizable in any dimensions $d>2$, and it is simple to realize that it is not possible to find a renormalizable neutral coupling for the electromagnetic field. In the standard model, these terms appear as low energy effective terms. The coupling $\mu_1$ is of the order $(e/ m_e)^4$, while $\mu_2$ is of order $e/m_N$ for a neutron and $e^3 \, m_l/M_Z^2$ for a neutrino, where $m_N$ is the neutron mass, and $m_l$ represents the charged lepton masses.  These neutral terms have characteristic scales $\mu_1^{-1/4}$ and $\mu_2^{-1}$ where the effective description breaks down. In the standard model these terms are generated by integrating out charged fields. In fact, the charged particles appear at smaller scales than $\mu_1^{-1/4}$ and $\mu_2^{-1}$, where the effective model predicts new physics to occur.   

It is then evident we run into problems for constructing a purely neutral electrodynamics from the perturbative point of view. The question we want to address is if the same conclusion can be reached at the non perturbative level: is there a UV complete ``neutral electrodynamics'', or conversely, does any UV complete interacting theory for the photon contain charges?   

To address this problem in full generality we need to make the statement that ``there are no charges'' more precise. Charged operators do not exist because they are not gauge-invariant, and the existence of charged particles requires to understand the spectrum of the theory. We choose a different route. The model (\ref{lagra}), taken at the classical level, has two different closed form fields
\bea
&& F_{\mu\nu}\,,\hspace{7.4cm} d\wedge F=0\,,  \\
&& G_{\mu\nu}=  \varepsilon_{\mu\nu\alpha\beta} \, (F^{\alpha\beta} (1+\mu_1 \, F^2) +\mu_2 \, \bar{\psi}\, \sigma^{\alpha\beta} \,\psi)\,,\qquad d\wedge G=0\,. 
\eea
 The field $G$ is closed due to the equations of motion. The fact that there are two closed two-form fields $F$ and $G$ is what we interpret as the absence of charges in the model. In fact, magnetic charges would lead to $d\wedge F\neq 0$ and electric charges to the impossibility to find a field $G$ such that $d\wedge G=0$. Writing the equation of motion in the usual way $\partial_\nu \,F^{\nu\mu} = J^\mu$  the only way to obtain a closed two form $G$ would be that 
\be
J^\mu\propto \partial_\nu \omega^{\mu\nu} \label{form}
\ee
 for some other gauge invariant form-field $\omega^{\mu\nu}$. Then  $G=*(F+\omega)$, where $*$ is the Hodge dual (contraction with the Levi-Civita tensor). The particular form  (\ref{form}) of the current is precisely what we get in the present model, but it is not the case for the usual quantum electrodynamics. Eq. (\ref{form}) means that there is a redefined non trivial electric flux operator given by the integral of $G$ on two dimensional surfaces which does not detect any charges through the Gauss law. 

To further clarify and generalize this situation, it is convenient to use the idea of generalized symmetries \cite{Gaiotto:2014kfa}, and specially an understanding of these symmetries in terms of Haag duality violations \cite{Casini:2020rgj,Casini:2021zgr}. We formulate the problem in detail in these terms in the next section. Here we anticipate 
that the effective model (\ref{lagra}) has two form-symmetries corresponding to the group $\mathbb{R}$. The first is just the group of ``non local'' operators $e^{i q \Phi_F}$, $q\in \mathbb{R}$, where $\Phi_F$ is the  flux of the closed form $F$ on a two dimensional (non closed) surface.  This is a set of Wilson loops running along the boundary of the surface and corresponding to charge $q$. Second, in a dual way we have the t' Hooft loop operators $e^{i g \Phi_G}$, $g\in \mathbb{R}$, that also form a group $\mathbb{R}$. We are writing $\Phi_G$ for the flux of second closed form field $G$ on a two dimensional surface. 
Then, we have a dual set of form-symmetries given by non-compact groups. If we now add to the model~(\ref{lagra}) a further (non gauge-invariant) field of electric charge $q_0$ we can form a gauge invariant Wilson line ended in these operators. The Wilson loop of the same  charge $q_0$ can then be decomposed into a product of Wilson lines, and it is not any more a non local operator along the path. 
The same will happen for operators of charge $n \, q_0$, $n\in \mathbb{Z}$. Therefore the group of non local classes of Wilson loop operators is reduced to a $U(1)$ group, with $q\in [0,q_0)$. When there are electric charges,  the electric fluxes are not conserved any more, and only operators of charge $g=2\pi m/q_0 $, with $m\in Z$, are genuine line operators violating Haag duality, rather than surface operators. The group of non local t' Hooft loops is then reduced to $\mathbb{Z}$. The dual of a compact group such as $U(1)$ is always a discrete group such as $\mathbb{Z}$. Therefore, the dual symmetry in this case is not continuous,  and cannot have an infinitesimal generator such as the flux of $G$. If there are electric charges we will not be able to find an exactly closed quantum field $G$. But of course $G$ could exist in an approximate effective way at low energies. The same happens in a dual way: if there are magnetic charges there is no such a symmetry generator as a closed field $F$. 

In these terms, we can rephrase what is special about any neutral electrodynamics model by saying that it possesses a non-compact form symmetry. This is reflected in the existence of the infinitesimal generators given by the two dual closed fields $F$ and $G$. In \cite{Benedetti:2022zbb} we conjectured that non-compact  generalized symmetries can only happen for free models. This is precisely the content of the present paper restricted to the case of generalized symmetries generated by local currents, namely form-symmetries.\footnote{We further comment on the more general case in the discussion section below.} We will show that $F$ and $G$, generators of the dual form-symmetries, can be chosen to be free dual fields. This means that if the effective theory fields $F,G$ are interacting, the generalized symmetry has to be broken down in the UV completion to a smaller group, implying the existence of electric and/or magnetic charges.               

Another example is the low energy theory of a Goldstone  boson. Such a model can only have derivative couplings
\be
{\cal L}= \frac{1}{2}\, (\partial \phi)^2+ \tilde{{\cal L}}(\partial \phi)\,.\label{lagra1}
\ee
In this case, the two dual forms are given by
\be
F_{\mu_1\cdots \mu_{d-1}}= \varepsilon_{\mu_1\cdots \mu_{d-1}\mu} \, \left(\partial^\mu \phi + \frac{\partial \tilde{{\cal L}}}{\partial (\partial_\mu \phi)}\right)\,, \qquad G_\mu= \partial_\mu \phi\,.
\ee
Both of these fields are closed $d\wedge F=d\wedge G=0$ and the effective theory has non-compact form-symmetries generated by $d-1$ and $1$-form fields. Again, all possible couplings are non renormalizable since the dimension of $\partial_\mu \phi$ is $d/2$ and must be accompanied in the Lagrangian by a vector field of dimension smaller than $d/2$, which is not possible. 
 The question is whether the theory can be completed retaining the non-compact form-symmetry. When this Lagrangian comes from a symmetry breaking of a compact Lie group, $F$ turns out to be the conserved current of the symmetry. As such it can persist as a well defined conserved field in the UV. On the other hand, since the field $\phi$ lives in a compact space, it is not a well defined field operator in the usual sense. Then, it does not follow that $\partial \phi$ is a closed $1$-form field exactly, and this will fail at energies larger than the symmetry braking scale. This is similar to the previous case of neutral electrodynamics when including electric charges. One current remains closed but it only generates a compact group. The dual generalized symmetry becomes discrete and it cannot be generated by a local field.

In appendix \ref{clasical} we show that models of neutral electrodynamics and effective theories of Golstone bosons do not have any intrinsic problem at the classical level. The notion of non-compact form-symmetry can be defined at the classical level and those models display those symmetries.  This allows to express the result of this paper with a slightly different perspective. As the generalized symmetries of the classical models cannot be implemented quantum mechanically, this can be interpreted as an anomaly. In the present case, this anomaly coincides with, and expresses in a more rigorous way, the idea of the non renormalizability of the concrete Lagrangian examples.   
 
The main observation behind the proof that the closed fields $F$ and $G$ are free and massless is quite simple and it only involves the analysis of the two point functions. We will see that the correlator $\langle F G\rangle$ is non zero, contains a term that obeys the massless equation of motion, and further it cannot renormalize. On the other hand, $\langle F F\rangle$ and $\langle G G\rangle$ can renormalize. By analysing the UV fix point it is possible to ``filter'' the free massless parts of these fields. 
 
The plan of the paper is as follows. In the next section we describe the setup and the properties of the two generators of dual non-compact form symmetries.  In section \ref{twopoint} we analyse the form of the most general two point functions of these fields, taking into account the symmetries (or antisymmetries) required and the positivity constraints. To this end we need to understand the  the two-point functions as biforms, so we review how some of their properties extend from the usual approach to differential forms and review our notation in appendix \ref{notation}. In section \ref{scale} we study aspects of generalized symmetries in scale invariant theories. In particular we prove that in such a case the closed form current fields that generate a non-compact form symmetry must be free and massless. Section \ref{filtering} contains the main proof. It extends the results obtained for scale invariant theories to all QFTs . This will require the study of the UV fix points of the theory, and specially the constraints they impose over the full QFT. Conveniently,   in appendix \ref{RGs} we review some recent results concerning the mathematical approach to RG flows that are relevant to this discussion.  Finally, section \ref{discussion} contains an open discussion on the results, and describes connections with other problems. We have already mentioned appendix \ref{clasical} containing a definition of non-compact symmetries for classical fields. Also, appendix \ref{free} contains a new proof that a field obeying a linear equation of motion is free.   Finally, appendices \ref{AppFlux} and \ref{fild2} contain some lengthy calculations that might be of use to the reader.

\section{General setup for non-compact form symmetries}
\label{setup}

Non-compact form-symmetries correspond to the existence of a closed real $k$-form field $F$, $d F=0$, and a closed $q$-form field $G$, $d G=0$, with $d=k+q$, and $1\le k,q \le d-1$. We use the short-cut notation $dF=d\wedge F$. For definiteness we take $k\ge q$.  
Both $F$ and $G$ are assumed to be physical (gauge invariant) Wightmann fields.

The fluxes 
\be
\Phi_F= \int_{\Sigma_F} F\,,\hspace{1cm}  \Phi_G= \int_{\Sigma_G} G\,, \label{fffg}
\ee
over spatial $k$-dimensional surfaces $\Sigma_F$, and $q$-dimensional surfaces $\Sigma_G$, depend only on the boundaries $\Gamma_F=\partial \Sigma_F$, $\Gamma_G=\partial \Sigma_G$, and not on the particular surfaces having these same boundaries. For simplicity, we can think that $\Sigma_F$ and $\Sigma_F$ lie at $x^0=0$.
If $k<d-1$, both $\Phi_F$ and $\Phi_G$ commute with any local field operator at any point $x$ spatially separated from $\Gamma_F$ and $\Gamma_G$ respectively. This is because the surfaces $\Sigma_F$ and $\Sigma_G$ can be deformed such as to be spatially separated from any such point $x$, without changing the flux operator. In particular, the fluxes commute with $F(x)$ and $G(x)$ for $x$ spatial to the boundaries. If $k=d-1$ the commutativity of  $\Phi_F$ with local operators spatially separated from the boundary cannot be implied because we can only  deform the surface $\Sigma_F$  in a time-like direction. In this case, we further assume $\Phi_F$ commutes with $F(x)$ and $G(x)$ for $x$ spatial to $\Gamma_F$. In this case $k=d-1$ the flux of $F$ over all space gives a generator of a global one parameter group of symmetries. Then, the additional requirement for this case is that $F$ and $G$ are uncharged fields with respect to this symmetry.  Note that, without loss of generality, we can take the Hilbert space as the one generated by $F,G$ acting on the vacuum, such that these fields act irreducibly.    

Take  $\Gamma_F$ and $\Gamma_G$ with the topology of $S^{k-1}$ and $S^{q-1}$, and simply laced to each other in the spatial plane $x^0=0$. Let them be the boundaries of disk like regions $\Sigma_F, \Sigma_G$, of dimension $k$ and $q$ respectively. A last assumption involved in the idea of non-compact form-symmetries is that the fluxes $\Phi_F$ and $\Phi_G$ do not commute to each other in this case of simply laced boundaries. This implies that none of the form fields is a physically exact form-field. That is, they cannot be written as $F=d  f$, or $G=d g$, for gauge invariant fields $f,g$.  Otherwise $\Phi_F$ or $\Phi_G$ could be written as integrals of local fields on $\Gamma_F$ and $\Gamma_G$ and the fluxes would commute.\footnote{An exception is the case of global symmetries $k=d-1$, $q=1$, where $G$ could be of the form $d \phi$, for a scalar field $\phi$. However, in order to have non commuting fluxes $\phi$ must be an operator charged under the global symmetry, and thus falls out of the neutral algebras generated by $G$ and $F$.} 

A first simplification in this  scenario is the following. The commutator $[\Phi_F,\Phi_G]$ does not change if we deform $\Gamma_F$ or $\Gamma_G$  keeping them spatially separated and simply laced. The reason is that the change in the flux $\Phi_G$ under such a deformation of $\Gamma_G$ is a flux on a surface in between the boundary $\Gamma_G$ and its deformation $\Gamma_{G}'$. Then, it is a flux of $G$ in a region spatially separated from $\Gamma_F$, and commutes with $\Phi_F$. The same happens deforming $\Gamma_F$. 

As a result, we can displace both surfaces together to infinity keeping them linked, and the commutator cannot change. This implies the commutator of the fluxes commutes with any local operator, and therefore is a number, which is also a topological invariant for the pair $\Gamma_F,\Gamma_G$. We can normalize it to be
\be
[\Phi_F,\Phi_G]=i\,.\label{fluxc}
\ee

 In this situation the theory generated by the fields $F,G$ contains violations of Haag duality, as described in \cite{Casini:2020rgj,Casini:2021zgr}. To see this more clearly, take causally complementary\footnote{Two space-time regions $R_1$, $R_2$, are causally complementary if $R_1'=R_2$, and $R_2'=R_1$, where $R'$ is the set of points spatially separated from $R$.} regions $R_F$ and $R_G$, such that $\Gamma_F\subset R_F$ and  $\Gamma_G\subset R_G$. Then, we can assign to these regions   von Neumann  algebras  ${\cal A}(R_F)$ and ${\cal A}(R_G)$  generated by both fields $F,G$ in $R_F$ and $R_G$. 
However, these algebras do not include the one parameter groups of unitaries $a(q)= e^{i q \tilde{\Phi}_F}$, $b(g)=e^{i g \tilde{\Phi}_G}$, $q,g\in \mathbb{R}$, formed by  exponentials of smeared versions $\tilde{\Phi}_F,\tilde{\Phi}_G$ of the fluxes, where the boundaries have support inside the respective regions.\footnote{For a longer detailed description of these smeared fluxes and their relation to Haag duality violation see \cite{Pedro,casini2021generalized}.} To see this note that by construction $a(q)$ commutes with ${\cal A}(R_G)$ and $b(g)$ commutes with ${\cal A}(R_F)$. Then the algebra ${\cal A}(R_F)$ cannot contain $a(q)$ because all the elements of this algebra commute with the flux $\Phi_G$ while $a(q)$ does not. In the same way, $b(g)$ is not contained in ${\cal A}(R_G)$.  So we have
\bea
{\cal A}(R_F)\subsetneq {\cal A}(R_G)'\,,\label{33}\\
{\cal A}(R_G)\subsetneq {\cal A}(R_F)' \,.\label{34}
\eea
Here ${\cal A}'$ means the algebra of all operators that commute with ${\cal A}$. Eqs. (\ref{33},\ref{34}) imply there is no Haag duality for these regions.\footnote{For two causally complementary regions $A'=B$, $B'=A$, causality implies that the corresponding algebras commute, ${\cal A}(A) \subseteq {\cal A}(B)'$, ${\cal A}(B)\subseteq {\cal A}(A)'$. Haag duality is a form of maximality for the local algebras in which ${\cal A}(B)={\cal A}(A)'$ and viceversa. } The operators $a(q)$, $b(g)$, are non local operators in their respective regions in the sense that they cannot be formed locally in these regions, but still commute with the local operators outside them. The form-fields generate the dual continuous groups of  non local symmetry operators in a way analogous to the way a Noether current generates a continuous global symmetry.   The charges $q,g,$ describe different non local classes of operators. Members of the same non local class differ by the action of local operators in the regions. The non local operators $a(q), b(g)$ are dual to each other in the sense they are based on complementary regions and do not commute with each other. Both sets of dual non local operators form continuous groups, and we take this scenario as a definition of a non-compact form symmetry.   

The relation (\ref{fluxc}) eliminates the possibility that any of the dual groups be a compact $U(1)$ group, and gives two dual non-compact $\mathbb{R}$ groups of generalized symmetries.\footnote{If one of the dual symmetries were $U(1)$ the dual one would be forced to be a non continuous $\mathbb{Z}$ group, and its non local operators could not be generated by a form-field in a way analogous in which a discrete global symmetry is not generated by a current. } The commutation relations for the non local operators is fixed to be 
\be
a(q) \, b(g)=e^{-i\, q\, g}\, b(g)\, a(q)\,.
\ee

\section{The two point functions}
\label{twopoint}
In this section we analyse the general form of the two point functions of the real form
fields $F$ and $G$ in any dimension. We start with the study of all the possible tensor structures in momentum space. Then, we provide the most general expression  for correlators in the Kallen-Lehmann representation allowed by conservation of the fields, spatial commutativity, and the flux commutator (\ref{fluxc}). The positivity constraints required on the spectral Kallen-Lehmann functions in each case are also  given.

\subsection{Tensor structures}
Any of the two point functions of $F$ and $G$ is a biform conserved in each index. In other words, in momentum space the generic structure of the Wightman function  is of the form  
\be
 \int \frac{d^{d}p}{(2 \pi)^{d-1}} \,\theta(p^0)\,\theta(p^2)\, e^{i p x}\, P_{\mu|\nu}(p)\,.\label{kl1w}
\ee
where $\mu,\nu$ in the polarization tensor $P_{\mu|\nu}(p)$  are antisymmetric $(k|k)$, $(q|q)$, $(k|q)$ or $(q|k)$ multi-indices, depending on the particular two point function.  See appendix \ref{notation} for a more extended review of the notation used thought this paper for biforms. The conservation $dF=0,dG=0,$ requires that 
\be
p\,\wedge P=P\wedge p\,=0\,, \label{con}
\ee
 where the notation means the wedge product acting on the left or right indices respectively.  Using the Hodge star operator $*$, we can analyse an equivalent dual problem. If we define  $\tilde{P}=* P *$, the conservation condition is given by $p\cdot\tilde {P}=\tilde{P}\cdot p=0$. This is because these two tensors are proportional 
\be
* p \wedge * (\cdots) \,\sim \, p\cdot (\cdots) \,. 
\ee
This corresponds in momentum space to the identity $\delta =(-1)^{k d+1} *\, d\, * $ between the coderivative $\delta$ (proportional to the divergence of the tensor)  and the exterior derivative  $d$ acting on forms. We remember that $\delta \,\delta=0$ as a consequence of $d \, d=0$. 

 In what follows all the possible structures for the polarization tensor $P$ are studied. This tensor must be written in terms of the metric $g$, the momentum $p$, and the Levi Civita tensor $\varepsilon$. First, we consider tensors constructed using  the metric alone . Let us start with the case of $\langle FF\rangle$ with $(k|k)$ biforms of equal number  of  indices. The case of $\langle GG\rangle$ follows in a like manner for $(q|q)$. Antisymmetry implies that if there is a metric tensor its two indices must belong to the two different sets of indices $\mu$, $\nu$, of equal size. Then, there are exactly $k$ metric tensors in $P$. After antisymmetrising on both sets of indices there is only one possible tensor structure of this form, given by a term proportional to the generalized Kronecker symbol
\be
g^{(k)}_{\mu|\nu}=\sum_\sigma \textrm{sgn}(\sigma)\, g_{\mu_1\nu_{\sigma(1)}}\cdots  g_{\mu_k\nu_{\sigma(k)}}=\frac{(-1)^{d-1}}{(d-k)!} \varepsilon_{\mu}^{\,\,\,\alpha} \varepsilon_{\nu\alpha }\,,\label{yu}
\ee 
where the sum is over the permutations $\sigma$ of the set $(1,\cdots,k)$, $\textrm{sgn}(\sigma)$ is the signature of the permutation, and $g_{\mu\nu}$ denotes the usual Minkowski metric in signature $(+--\cdots)$ .
 For a polynomial of the momentum and the metric, by antisymmetry, we cannot include more than two powers of $p$, nor just one momentum, because that would imply one of the metrics has the two indices in the same antisymmetric set. 
 With two momentum vectors and the metric the only possible tensor structure contains a momentum on each set of indices  
\be
p\cdot g^{(k+1)}\cdot p \label{ten}\,.
\ee     
Nevertheless, only one combination of (\ref{yu}) and (\ref{ten}) is closed in both indices. Since (\ref{ten}) is the only bi-coclosed structure, the bi-closed  one for $P$ is its dual, for the appropriate order, 
\be
* p\cdot g^{(d-k+1)}\cdot p\, * \,  \sim\, p\wedge g^{(k-1)}\wedge p\,. \label{340}
\ee 
Now consider the correlator $\langle F G\rangle$, with $(k|q)$ indices, $k+q=d$. The same analysis applies if $k=q=d/2$ for $d$ even. If $k\neq q$ nothing antisymmetric can be formed with the metric alone. With one momentum we have for $k=q+1$, $d=2 k-1$ odd, 
\be
g^{(k)}\cdot p\sim p\wedge g^{(k-1)}\,.\label{35}
\ee
 This is however only closed on one side but co-closed on the other, and cannot appear in the correlator.
 
 Furthermore, with two or more momentum vectors, antisymmetry again does not allow any tensor structure constructed with the metric and the momentum alone for $k\neq q$.  However, we can consider  tensor structures containing the Levi-Civita tensor $\varepsilon$. These result from the previous ones by acting with the Hodge star in one side only. The part of $P_{\mu\nu}$ containing $\varepsilon$ has to be separately bi-closed. We then have to go through the previous tensor structures and check whether they are closed on one side and co-closed on the other.  The only possible bi-closed structure  is obtained by dualising (\ref{340}) when $p^2=0$, as in such a case we get
\be
* p\cdot g^{(d-k+1)} \cdot p\sim p\cdot g^{(k+1)}\cdot p *\,.\label{3100}
\ee 
This can always appear in $\langle F G\rangle$. It can also appear in $\langle F F\rangle$ or  $\langle G G\rangle$  when $k=q=d/2$.  
 
When including the Levi-Civita tensor, a new posibility linear in the momentum emerges for the correlator $\langle F F\rangle$. This is recovered by dualising (\ref{35})  on one side 
\be
* g^{(d-k)} \wedge p\sim p\wedge g^{(k-1)} *\sim *\, p\cdot \varepsilon *\,. \label{317}
\ee 
This can only appear when $d$ is odd and $d=2 k-1$. In $d=3$ represents  the case of the Maxwell-Chern-Simons field.  
  
In summary, for the generic case where $k\neq q$ and $k\neq (d+1)/2$, the  most general form of the correlators in Kallen-Lehmann form is
\bea 
\langle F(x) F(0)\rangle &=& \int_0^\infty ds\,( a_F\, \delta(s) + \rho_F(s))\,\int  \frac{d^{d}p}{(2 \pi)^{d-1}} \,\theta(p^0)\,\delta(p^2-s)\, e^{i p x}\,\, P^{(k)}(p)\,,\label{pri}\\
\langle G(x) G(0)\rangle &=& \int_0^\infty ds\, (a_G\,\delta(s)+\rho_G(s))\,\int  \frac{d^{d}p}{(2 \pi)^{d-1}} \,\theta(p^0)\,\delta(p^2-s)\, e^{i p x}\, P^{(q)}(p)\,,\label{pri2}\\
\langle F(x) G(0)\rangle &=& \,\int  \frac{d^{d}p}{(2 \pi)^{d-1}} \,\theta(p^0)\,\delta(p^2)\, e^{i p x} \, (P^{(k)} \tilde{*})(p) \,,\label{kkj1}\\
\langle G(x) F(0)\rangle &=& \,\int  \frac{d^{d}p}{(2 \pi)^{d-1}} \,\theta(p^0)\,\delta(p^2)\, e^{i p x} \, (* P^{(k)})(p) \,,\label{ppri}
\eea
where  the operator $\tilde{*}$ is defined from the original Hodge dual as $\tilde{*}= (-1)^{kq} *$. The Kallen-Lehmann functions have to be real because of commutativity at spatial distances and the fact that fields are real. Here, we have singled out the massless parts of the spectral densities for $\langle FF\rangle $ and $\langle GG\rangle$ with the real coefficients $a_F$ and $a_G$. Also, we have chosen the normalization of (\ref{340}) as
\be
P^{(k)}_{\mu|\nu}\equiv\frac{(-1)^{k-1}}{(k-1)!}\,g_{\,\,\,\,\mu|\alpha}^{(k)\,\,\,\,\gamma}\, g^{(k)}_{\nu|\beta\gamma}\,\, p^\alpha\, p^\beta\,.\label{jj1}
\ee   
This allows us to obtain a useful identity for the dual of $P^{(k)}$ at both sides
\be
* P^{(k)} \tilde{*}= P^{(q)} +\,(-1)^{q}\,\, g^{(q)}\,\, p^2\,. \label{Pkq}
\ee 

 Interestingly, the correlator  $\langle F(x) G(0)\rangle$ can only be proportional to a unique term in momentum space  with support $p^2=0$, eq. (\ref{3100}), which we have written in the form (\ref{kkj1}) . This cannot be zero because the flux commutator has to be a number, and its vacuum expectation value does not vanish. We have chosen to normalize the coefficient to one.  
This correlator can only have a massless contribution, and therefore it is scale covariant, and satisfies the massless equation of motion 
\be
\square \,\langle F(x) G(0)\rangle=0\,.\label{ddos}
\ee 
The non-compact form symmetry forces the particular form of this two-point function that cannot renormalize. The fact that it obeys the free massless equation of motion contains the essence of the proof that the theory has a free massless sector. It is well known that a field $\phi$ whose two point function with itself satisfies the Klein Gordon equation also satisfies the equation at the operator level, $\square \phi(x)=0$. The linear equation of motion for the field implies the field is free. See below and appendix \ref{free}. However, the Klein Gordon equation for the correlator of two different fields such as (\ref{ddos}) does not imply the operator equation, nor that the fields are free. For example, we could write $F=F_0+F_1$ and $G=G_0+G_1$, with $F_0=* \,G_0$ free, and $F_1, G_1$  having zero two point functions between themselves and with the free fields. In this case we still have a free sub-sector of the theory, which is responsible for the non zero mixed two point function, and the non-compact form-symmetry. It is a natural expectation that this is the case in general. In the following we deal with the ``filtering'' of the fields to obtain their free part.    

As mentioned before, there are two special cases to consider. When $k=q=d/2$ the structures of $\langle FF\rangle$ and  $\langle GG\rangle$ can appear in $\langle FG\rangle$ and $\langle GF\rangle$ and viceversa. We write the new form of the correlators as in (\ref{pri}-\ref{jj1}) plus new terms
 \begin{align}
\langle F(x) F(0)\rangle & = \cdots + b_F\,\int  \frac{d^{d}p}{(2 \pi)^{d-1}} \,\theta(p^0)\,\delta(p^2)\, e^{i p x} \, ( P^{(k)}\tilde{*})(p)\,,\label{priaa}\\
\langle G(x) G(0)\rangle & = \cdots + b_G\, \int  \frac{d^{d}p}{(2 \pi)^{d-1}} \,\theta(p^0)\,\delta(p^2)\, e^{i p x} \, ( P^{(k)}\tilde{*})(p)\,,\\
\langle F(x) G(0)\rangle & = \cdots+ \int_0^\infty ds\,(c\, \delta(s) + \tilde{\rho}(s))\,\int  \frac{d^{d}p}{(2 \pi)^{d-1}} \,\theta(p^0)\,\delta(p^2-s)\, e^{i p x}\,\, P^{(k)}(p) \,,\label{kkj1aa}\\
\langle G(x) F(0)\rangle & = \cdots+ \int_0^\infty ds\,(c\, \delta(s) + \tilde{\rho}(s))\,\int  \frac{d^{d}p}{(2 \pi)^{d-1}} \,\theta(p^0)\,\delta(p^2-s)\, e^{i p x}\,\, P^{(k)}(p) \,.\label{pribb}
 \end{align}
The new term in the $\langle FF\rangle$  and $\langle GG\rangle$ correlators obeys that $ [P^{(k)}\tilde{*}]_{\mu|\nu}=(-1)^{k-1}[P^{(k)}\tilde{*}]_{\nu|\mu}$. Combining this with spatial conmmutativity  requires that  $ b_F, b_G =0$ if $k$ is even. In the other cases it just imply that the Kallen-Lehmann functions are real.

Similarly, from (\ref{317}), in the case of $d=2 k-1$ there is an additional possibility for the two-point function $\langle FF \rangle$ 
\be
\langle F(x) F(0)\rangle = \cdots +i^{k-1}\, \int_0^\infty ds\, \rho_{CS}(s)\,\int  \frac{d^{d}p}{(2 \pi)^{d-1}} \,\theta(p^0)\,\delta(p^2-s)\, e^{i p x} \, (*\, p.\varepsilon\, \tilde{*} )\,.\label{cs}
\ee 
where the factor $i^{k-1}$ has been added to keep $\rho_{CS}(s)$ real.
\subsection{Flux commutators}
As described above the linked flux commutators (\ref{fluxc}) are numerical and do not depend on the geometric form of the linked loops $\Gamma_F$, $\Gamma_G$. This can be verified by using the crossed correlator (\ref{kkj1}) in the expression for the flux commutator and integrating over the regions bounded by the loops $\Gamma_F$, $\Gamma_G$. See App.~(\ref{AppFlux}) for a explicit derivation.
 
The case $k=q=d/2$ may contain additional terms in the mixed two point function (\ref{kkj1aa}). This new term cannot change the flux commutator though. The reason is that the form of $P^{(k)}$ is proportional to (\ref{340}) and this implies these new terms\footnote{Physically, these type of terms correspond to the correlation between additive line operators, which, whether linked or not, ought to commute due to microcausality.} are the double exterior derivative (in different coordinates) of a $(k-1|q-1)$ biform that we call $ K(x,y)$  
\be
\langle F(x) G(y)\rangle = \cdots+ d_x \,d_y \,K(x,y)\,.\label{hh}
\ee
 We can see these doubly exact terms cannot change the value of the commutator in two  ways. First we can insert them directly in the expression for the flux commutator and verify its vanishing even when the fluxes are non trivially linked. See App.~(\ref{AppFlux}) for an example of this calculation. Another way is to change to the Euclidean correlators. These are non singular except at the coincidence points.  Take a $k$-dimensional disk $\Sigma_F$ with boundary $\Gamma_F$ at $x^0=0$ and a $q$-dimensional disk $\Sigma_G$ at $y^0=0$ with boundary $\Gamma_G$. The boundaries $\Gamma_F$ and $\Gamma_G$ are simply linked to each other.  
Without changes the fluxes we can deform $\Sigma_F$ moving the points to the future in Euclidean time such as to form a new surface $\Sigma_F^+$ with the same boundary $\Gamma_F$ at $x^0=0$. In an analogous way we form $\Sigma_F^-$ deforming the surface to the past. Let the time like Euclidean vector be $\hat{\tau}=(1,0.\cdots,0)$. The euclidean expression for the expectation value of the commutator is 
\be
\langle [\Phi_F,\Phi_G]\rangle=\lim_{\epsilon\rightarrow 0}\left(\int_{\Sigma_{F}^+ +\epsilon \,\hat{\tau}}  \int_{\Sigma_G} \langle F(x) G(y)\rangle  -  \int_{\Sigma_F^--\epsilon \,\hat{\tau}} \int_{\Sigma_G} \langle F(x) G(y)\rangle \right)\,.
\label{commm}
\ee
 It is immediate that the contribution of a doubly exact term like the one in (\ref{hh}) vanishes.  This is because, since this is exact in $x$ and $y$, we get an expression where the integration is on the boundaries $\Gamma_G$ and $\Gamma_{F}\pm \epsilon \hat{\tau}$. These two terms are continuous in the $\epsilon\rightarrow 0$ limit since only involve distant correlations, and cancel each other in the limit.  
This is not the case for the term (\ref{kkj1}) which is exact in any of its variables but not on both at the same time. This term can be integrated over $\Sigma_G$ to give an harmonic non exact form in $\mathbb{R}^d-\Gamma_G$ which contributes non trivially to the commutator (\ref{commm}). 

Viewed in the light of the euclidean calculation (\ref{commm}), the commutator of fluxes appears as a topological invariant for two intersecting surfaces of dimensions $k$ and $q$, $k+q=d$, one closed, $\Sigma_F^+ \cup \Sigma_F^-$, and another open $\Sigma_G$. This is called the Kronecker index for the surfaces, and the $\langle F G\rangle$ correlator is the ``linking number'' bi-form that allows to write this topological invariant as a double integral. See \cite{de1984differentiable}, chapter 33.

\subsection{Positivity} 
In a unitary theory, the  correlators of the form $\langle FF\rangle$ must be positive semidefinite 
\be
\int d^dx\,d^dy\, \phi^*(x)\, \langle F(x) F(y)\rangle\, \phi(y)\ge 0\,. 
\ee
This must be valid for all possible test functions $\phi(x)$. Therefore for each $p$ we have that
\be
\hat{\phi}^*(p) \, P^{(k)}(p)\, \hat{\phi}(p)\ge 0\, \label{smep},
\ee
which leads to the tensor structure in momentum space to be positive semidefinite. This 
 is true for  $P^{(k)}_{\mu\nu}$ as defined in (\ref{jj1}). It can be easily checked by  setting  
 $p$ in the time direction. For $p^2=0$ positivity of this matrix follows because it is the limit of a positive semidefinite matrix. 

Moreover, the positivity constraints apply separately to the massless delta functions and the remaning Kallen-Lehmann measures $\rho_F(s)$ and $\rho_G(s)$ appearing in (\ref{pri}) and (\ref{pri2}).  By choosing a smearing in (\ref{smep}) with $p^2\neq 0$, the positivity of the correlators  $\langle FF\rangle$ and $\langle GG\rangle$ imply $\rho_F(s)$ and $\rho_G(s)$ are positive measures in $[0,\infty)$.  Further, for tempered distributions $\rho_F(s), \rho_G(s)$ have to be at most of a polynomial increase at infinity. The positivity of the terms including a massless delta functions follows from positivity in the IR limit where these make the only remaining contribution. In this context, we recover that the constants multiplying the delta functions  must obey $a_F,a_G\ge 0$.

For this particular case, the positivity of the combined correlator matrix adds a new constraint only for $p^2=0$. We have conveniently chosen the normalization of (\ref{jj1}) so that $P^{(q)}(p)=* \,P^{(k)}(p) \,\tilde{*}$  when $p^2=0\,$. See eq. (\ref{Pkq}). Therefore, the correlator matrix for such case  writes
\be
\left(\begin{array}{cc}
a_F\,P^{(k)} & P^{(k)} \,\tilde{*} \\
* \, P^{(k)} &  a_G\, * \, P^{(k)} \,\tilde{*}
\end{array}\right)\,. \label{cm}
\ee
The requirement is that (\ref{cm}) should be a positive semidefinite  matrix. Then, it follows that positivity for $p^2=0$ gives
\be
a_F\, a_G\ge 1\,. \label{ines}
\ee
In particular, this further restrict  $a_F$ and $a_G$ to be non-zero, meaning that $a_F,a_G> 0$. This result, combined with the specific form of the correlators  (\ref{pri}) and (\ref{pri2}), imply that the theory has a massless particle.

We are left to analyze the special cases. When $k=q=d/2$  the correlators have the additional terms described in (\ref{priaa}-\ref{pribb}). For the massive part, we still have $\rho_F, \rho_G\ge 0$. For the $p^2=0$ sector, if $k$ is even, we get $a_F,\,a_G\,\geq0$ and  $a_Fa_G \geq 1+c^2$ . On the other hand, if $k$ is odd we have that $0\leq|b_F|\leq a_F$ and $0\leq |b_G|\leq a_G$, and the combined correlator matrix shields that $a_Fa_G + b_F b_G\geq 1+c^2$. Note that the combination of both restrictions in each case imply  the existence of a massless particle.

In the case $d=2 k-1$, the two-point function $\langle GG\rangle$ remains unchanged and we still recover $\rho_G(s)\ge 0$. However,  for $\langle FF\rangle$ defined by  (\ref{cs}),  we can check that
\be
0\leq |\rho_{CS}(s)|\leq \sqrt{s}\,\rho_F(s)\,.
\ee
This forces $\rho_{CS}(s)$  to vanish for $s= 0$ and does not alter the inequality (\ref{ines}) or the results obtained for the $p^2=0$ sector.

\section{Scale invariant form-symmetries  }
\label{scale}
We now analyse form-symmetries  in scale invariant theories. We start by showing that if the form-symmetry is non-compact, then the fields generating such a symmetry must be free and massless. Then, we study the more general case of a  single conserved form field, and classify different possibilities.

\subsection{Non-compact form-symmetries }
\label{scale1}

Consider the case of the non-compact form symmetry generated by the form fields $F$ and $G$ in a scale invariant theory. Let the dimensions of $F$ and $G$ be $\Delta_F$ and $\Delta_G$ respectively. This is achieved by spectral densities 
\be
\rho_F(s)\sim s^{\Delta_F-d/2-1}\,,\qquad  \rho_G(s)\sim s^{\Delta_G-d/2-1}\,.
\ee
The fact that these must be integrable measures implies the unitarity bounds 
\be
\Delta_F,\Delta_G\ge d/2\,.
\ee
 In the specific case of dimension $d/2$ saturating the unitarity bound,  the spectral measure has to be replaced by $\delta(s)$ instead of $s^{-1}$ because this latter is non integrable at $s=0$. On the other hand, the correlators (\ref{kkj1}-\ref{ppri}) require that
 \be
 \Delta_F+\Delta_G=d\,. \label{ppp} 
\ee
This gives $\Delta_F=\Delta_G=d/2$, saturating the unitarity bound, and having spectral measure $\delta(s)$. Then, the two point functions satisfy $\square_x \langle F(x) F(y)\rangle=0$. From here, by the standard argument that  
\be 
\langle \square F(x)\vert \square F(y)\rangle=0\rightarrow  \square F(x)|0\rangle=0\rightarrow \square F(x)=0\,, 
\ee
we get free equations of motion for the field (see e.g. \cite{Federbush:1960}). As it is well known this implies the field is free. There are several proof of this fact.  The references can be found in the appendix \ref{free} where we also give a simple alternative proof based on properties of harmonic functions. The fields can be normalized such as $F=* \, G$.\footnote{The parity odd term in $d=2 k-1$ cannot appear at the fix point, and the power counting for the case $k=q=d/2$ is not altered.}

Another way to arrive at this conclusion is to use a theorem by Buchholz and Fredenhagen  which implies that for a scale invariant theory a massless particle is a free particle \cite{buchholz1977dilations}. It is precisely the free particle content of the theory the responsible for the $\delta(s)$ terms in the spectral functions, in particular for the $\langle FG \rangle$ function that is necessary for the non-compact form-symmetry.

\subsection{A continuous form-symmetry in the scale invariant case }
\label{scale2}

In this section we analyse the more general case of a single scale invariant (continuous) form-symmetry generated by the form field $H$ with $h$ indices. We use another letter for this field to not confuse it with the more general analysis with $F,G$ above and below. This analysis not strictly necessary for the purposes of this paper, mainly focused on the non compact case, but we want to highlight that even if only scale invariance is invoked many of the features that follow from conformal invariance appear from the analysis of the generalized symmetry.   

 It turns out that the scaling dimension $\Delta_H$ of this field must be either $d/2$ or $h$ for $h\ge d/2$. 
The reason is quite simple. If $\Delta_H\neq h$ the non local operators $e^{i \alpha \,\int_\Sigma d\sigma H}$ have a parameter $\alpha$ that scales non trivially with scaling transformations. This means the non local classes of these operators (the charge of the form-symmetry labeled by $\alpha$) change under scaling. As the classes are non invariant under a continuous symmetry group, as shown in \cite{Benedetti:2022zbb}, the form symmetry must be non-compact. 
 Then, there is a continuous dual form-symmetry, say generated by $\tilde{H}$ with $d-h$ indices. As we have seen, the only form that the correlator of $H$ and $\tilde{H}$ has the necessary term to produce constant flux commutators is that both fields have the free dimension $d/2$. Another way to say this is that the flux operators $\Phi_H, \Phi_{\tilde{H}}$, generating the form-symmetry transformations, must have commutator $i$. The scale invariant transformation of parameter $\lambda$ multiplies one of the fluxes by $\lambda^{\Delta_H-h}$ and then the other must transform with exponent $h-\Delta_H$. This gives a scaling dimension $\Delta_{\tilde{H}}=d-\Delta_H$, and one of the two is incompatible with the unitarity bound except in the free case, where both are equal to $d/2$. Then, we also have the possibility of $\Delta_H=h$, for $h> d/2$. In this case, the form-symmetry is invariant under scaling, but it cannot be a non-compact symmetry as there cannot be a dual form with dimension $d-h<d/2$. It must then be a compact $U(1)$ symmetry.       

To summarize, a closed form-field $H$ in a scale invariant theory must be in one of the following mutually exclusive possibilities:\footnote{We are assuming an additive net can be defined for causal regions based on the $t=0$ surface, such that the form-symmetry can be properly defined. This explicitly eliminates generalized free fields. It is enough the theory contains a stress tensor, but could hold more generally.} 
\begin{enumerate}
\item[a)] H is a free field with dimension $\Delta_H=d/2$, generating a non-compact form symmetry together with its corresponding dual field $*\, H$.
\item[b)] H has dimension $\Delta_H=h$ with $h>d/2$, generating a continuous but compact $U(1)$ form symmetry.
\item[c)] H is an exact form field, namely, it is a total derivative $H= d \phi$ with $\phi$ a gauge invariant field, not generating a form symmetry. 
\end{enumerate}
In the last case the field is closed, but it does not produce a form-symmetry because the fluxes are local on the boundary. Still, the case $h=1$ can be a form-symmetry in the case where the dual symmetry is a global symmetry and the symmetry transformation acts as $\phi\rightarrow \phi+\textrm{const}$ for the scalar field $\phi$. This can only be a free massless scalar field $\phi$, and it is already covered in the case a).    

The previous reasoning did not involve conformal symmetry, but only arguments about Haag duality violation that imply the existence of dual generalized symmetries. However, it is interesting to analyze the implications of conformal invariance without appealing to the idea of Haag duality defects.\footnote{See \cite{Argyres:1995xn,hofman2019goldstone} for similar analysis in the conformal case.}
 
A primary $h$-form field $H$ has unitarity bound $\Delta_H\ge \max (h, d-h)$ \cite{mack1977all,siegel1989all,minwalla1998restrictions,costa2015conformal}. 
 It is closed only for $\Delta_H=h\ge d/2$, when it saturates the unitarity bound. Conversely, it is co-closed for $\Delta_H=d-h$, $h\le d/2$. 
The only free case is for even $d$ with $h=d/2$. 
 Out of these cases, the closed field cannot be primary. 
 In this case, we have to analyse if it can be a descendant field, that is, a derivative of a primary field. A derivative can add an index to the field, in which case only one derivative is allowed because the fields are antisymmetric. That is, we should have $H=d \phi$. Another possibility would be that $H_\mu=\partial^{\alpha_1}\cdots \partial^{\alpha_n} \phi_{\mu \alpha_1\cdots \alpha_n}$ for a primary field $\phi_{\mu \alpha_1\cdots \alpha_n}$ antisymmetric in the indices $\mu$ and symmetric in the $\alpha$'s. However, this cannot be closed because such primary fields do not obey conservation equations unless at the unitarity bound, and in such a case the divergence is zero \cite{costa2015conformal}.
The third and last possibility would be that $H=\delta \phi$ for an antisymmetric primary $\phi$. Conservation implies $\phi$ is free, and that implies the number of indices in $\phi$ is $d/2$, and that $\delta \phi=0$. Therefore for a CFT the result is the same as above, with the addition that case a) can only happen for $h=d/2$.  Summarizing, if we have a conformal  fix point with a non-compact form symmetry, it can only correspond to the theory of two free primary forms of dimension $d/2$ (as in the Maxwell field for $d=4$) or the free Goldstone boson case.

Returning to the general case, it is important that the form-symmetries of cases a) and b) happen not to be saturated. Equivalently, we cannot have arrived at a topological limit in which unitary non local operators have expectation value $1$ or $0$. These cases can only be the result of a limit, but never be produced by actual operators existing in the theory.  For example, take a region $R$ with the topology of a $S^{k-1}\times T$ with $T$ being a compact subset of $\mathbb{R}^{d-k+1}$. Then, we can form the unitary operator  
 \be
 W(q)=  e^{i \, q \,\int \,\omega(x)\,H(x)}\,, \label{45}
\ee
where $q$ is the charge of the non local class, $\omega(x)$ is a smearing function such that 
\be
\delta \omega(x)= J(x)\,, \qquad J(x)=0 \,\,\,\text{if}\,\,\, x\notin R\,.\label{46}
\ee
This warrants that $W(q)$ is an operator localized in $R$ in the sense that it commutes with local operators outside of $R$. To have charge $q$ we need to normalize the flux of $J$ on a section $S$ as of $R$
\be
\int_S J=1\,. \label{47}
\ee
The expectation value of this unitary operator is bounded as  $0< |\langle 0| W(q)|0\rangle| <1$. It cannot be equal to $1$ because in that case $W(q)|0\rangle=|0\rangle$ by Cauchy Schwartz inequality. By Reeh-Schlieder theorem if  $(W(q)-1)|0\rangle=0$ for a local operator then $W(q)=1$, which is not the case. In the same way it cannot be $\langle 0| W(q)|0\rangle=0$ for all $q$. If that where the case, as $W(q)$ cannot annihilate the vacuum, it must convert it into an orthogonal unit vector $|q\rangle$. It would follow that all the $|q\rangle$ are orthogonal to each other for the continuous parameter $q$, what is impossible.    

This has an interesting implication in any theory in which there is a closed field $F$ (not necessarily associated with any non-compact symmetry). In that case, either the field is a total derivative $F=d \phi$, or the UV fix point contains Haag duality defects. That is, Haag duality defects cannot become topological at the UV. This is in contrast to the cases of non continuous sectors, as the case of asymptotically free Yang Mills theories where the discrete generalized symmetry becomes saturated at the UV \cite{casini2022entropic}.

\section{Renormalization group flow }
\label{filtering}

In this section we prove that a theory with non-compact form-symmetries has a massless free sector. Such results descend from the scale invariant proof presented in section \ref{scale1} via arguments concerning the renormalization group flow.  Therefore, to proceed, we will be making the usual assumption that a UV complete theory has a UV scale invariant fix point, and that the full theory arises from the UV fix point by perturbing it away from criticality. For our interest in the present problem, in fact, we need to assume less structure, basically that there is a UV scale invariant fix point, and that to quantum fields in this fix point corresponds quantum (Wightman) fields in the complete theory, and viceversa. As the details of the expected relation between the UV fix point and the QFT are rarely spelled out, we endeavor to be more precise in what follows. In this scenario we  analyse constraints over the full QFT arising from the existence of a UV fix point with a non-compact form-symmetry. Though the existence of a scale invariant fix point is an assumption in this paper, we note that this has been proved under the condition of a phase space property that restricts the increase of the number of degrees of freedom at high energies \cite{bostelmann2010dilation}.

\subsection{Assumptions about the RG flow and the UV limit}

Since the existence of a completion, or a  UV limit of the theory, is quite central to our arguments, we are going to be explicit about the assumptions involved in this idea. We describe the QFT and its UV fix point by the collection of their Wightman fields. We assume the UV fix point is a scale invariant theory. 
 
Let $\varphi$ be a field. We assume there is always a $\Delta>0$ such that\footnote{See the remark about this point on appendix \ref{RGs}.}
\begin{align}
&\lim_{\lambda\rightarrow 0} \lambda^{\alpha} \langle \varphi(\lambda x) \varphi(0)\rangle= 0\,, \qquad\quad \forall\,\, \alpha> \Delta\,, \label{primo}\\
& \lim_{\lambda\rightarrow 0} \lambda^{\alpha} |\langle \varphi(\lambda x) \varphi(0)\rangle|= \infty\,,\qquad \forall\,\, \alpha< \Delta\,.\label{sec}
\end{align}
Then for such a field $\varphi$ we say it has asymptotic dimension $\Delta$. In a scale invariant theory this coincides with the scaling dimension for irreducible fields. We assume, both for the UV fix point and the QFT, that the linear space of fields with dimension less than any $\Delta$ is finite dimensional and that the fields are in finite dimensional representations of the Lorentz group. This is a necessary condition for  many usual requirements of a QFT, for example having a finite partition function.

To proceed, regarding the relation between the full QFT and its UV fix point we will be making the following  assumptions:

\noindent \textbf{1)\,\,}  For each $\varphi$ of the QFT there is a function $Z_\varphi(\lambda)$ and a  field $\varphi_0$ of the UV fix point (we call $\varphi_0$ the UV limit of $\varphi$), unique up to normalization, such that 
\be
\lim_{\lambda\rightarrow 0}\,\langle Z_{\varphi}(\lambda)\,  \varphi(\lambda x_1)\cdots Z_{\varphi}(\lambda)\varphi(\lambda x_n)\rangle =  \langle \varphi_0( x_1)\cdots \varphi_0(x_n)\rangle \,.\label{eq}
\ee
The functions $Z_\varphi(\lambda)$ are highly non unique but their asymptotic limit is quite restricted. In particular 
\begin{align}
& \lim_{\lambda\rightarrow 0} \lambda^{-\alpha}  \, Z_{\varphi}(\lambda)=0\,, \qquad\,\,\,\text{if}\,\, \alpha<\Delta\,, \label{una} \\
& \lim_{\lambda\rightarrow 0} \lambda^{-\alpha}  \, Z_{\varphi}(\lambda)=\infty\,, \qquad\text{if}\,\, \alpha> \Delta\,, \label{dos}
\end{align}
where $\Delta$ is the scaling dimension of $\varphi_0$.
Form this it follows that $\varphi_0$ has a unique scaling dimension and spin representation which coincide with the asymptotic dimension and spin of 
 $\varphi$.  We write this field mapping (up to normalization) ${\cal M}(\varphi)=\varphi_0$, or simply $\varphi\rightarrow \varphi_0$.
 This mapping is generally many to one because from a linear combination of fields in the QFT, only the highest dimension component survives in the limit.  It also follows that for an irreducible non zero combination of derivatives of $\varphi$,  that for short we write $\partial \varphi$, and such that $\partial \varphi_0\neq 0$, we have ${\cal M}(\partial \varphi)=\partial \varphi_0$, where $Z_{\partial\varphi}(\lambda)= \lambda\, Z_\varphi(\lambda)$. 
 If $\partial \varphi=0$ the same holds for $\varphi_0$, namely $\partial\varphi_0=0$. 

\noindent \textbf{2)\,\,}   There is a linear basis $B_0$ of the fields at the UV theory, having definite spin and scaling dimension, and a basis $B$ for the fields of the QFT, such that for each of the fields $\varphi_0\in B_0$ there is a unique (up to normalization) ${\cal N}(\varphi_0)=\varphi\in B$, such that ${\cal M}(\varphi)={\cal M}({\cal N}(\varphi_0))=\varphi_0$. The physical idea behind this assumption is that irreducible fields in the UV fix point (in a certain basis in the case of degenerate spectrum) generate fields in the QFT once the theory is perturbed away from the fix point.  
 From this it follows that for each $\tilde{\Delta}>0$, the fields in ${\cal N}(\{\varphi_0 \in B_0, \Delta_{\varphi_0}< \tilde{\Delta}\})$ form a linear basis for all fields in the QFT with asymptotic dimension less than $\Delta$, and the dimensions of the two spaces is the same. It is expected that any $\varphi\rightarrow \varphi_0$ leads to a $\varphi_0\in B_0$, excepting degeneracies due to symmetries. More generally, for such $\varphi\rightarrow \varphi_0$, where $\varphi_0$ is decomposed linearly into a subset $\varphi_0^i \in B_0$, the field $\varphi$ can be decomposed linearly into the elements $\phi^i$ of $B$ associated to $\varphi_0^i$ plus, eventually, fields with lower or equal dimension.   Combining with the first assumption, we ask that correlation functions obey
 \be
\lim_{\lambda\rightarrow 0} \langle Z_{\varphi^{i_1}}(\lambda)\,\varphi^{i_1}(\lambda x_1)\cdots Z_{\varphi^{i_n}}(\lambda)\varphi^{i_n}(\lambda x_n)\rangle =  \langle \varphi^{i_1}_0( x_1) \cdots \varphi^{i_n}_0(x_n)\rangle \,.\label{eq2}
\ee

\noindent  \textbf{3)\,\,} For any $\varphi_0\in B$, with ${\cal N}(\varphi_0)=\varphi$, and any non zero irreducible field formed from the derivatives of $\varphi_0$, that for short we call $\partial \varphi_0$,  we assume that $\partial \varphi_0 \in B_0$ and ${\cal N}(\partial \varphi_0)=\partial \varphi \in B$. 

 Though these are assumptions implicit in the idea of a UV limit theory, it would be important to have a derivation of these properties from a more general standpoint. Though we will not deal with this investigation here, we note that 
progress in mathematical studies of QFT in the last decades help to delineate the contours of this standard lore assumptions. For the benefit of the reader, we briefly review part of this progress and relevant references in appendix \ref{RGs}.  


\subsection{Filtering at the UV }

Let us analyse the different possibilities for the UV limits $F_0, G_0$ of the fields $F,G$. These satisfy $d F_0=d G_0=0$. Let us start from the simplest possible scenario in which the form-symmetry is generated at the UV by the limit fields $F_0, G_0$. As discussed in section \ref{scale} this implies both fields have dimension $d/2$. The correlations of these UV fields will be proportional to the $\delta (p^2)$ in momentum space. The requirement that the linking term in $\langle F G\rangle$ is not erased by the renormalization implies that the renormalization functions have a  limit $Z_F\sim \lambda^{d/2} \, Z_F^0$, $Z_G\sim \lambda^{d/2}\, Z_G^0$, with finite $Z_F^0, Z_G^0$. Using the notation of (\ref{pri}-\ref{ppri}), we get for the coefficients of the normalized tensor structures for $p^2=0$ in the matrix correlator of $F_0,G_0$:
\be 
\left( \begin{array}{cc}
(Z_F^0)^2\, (a_F+\int ds\, \rho_F) & Z_F^0\, Z_G^0    \\
Z_F^0\, Z_G^0  & (Z_G^0)^2\, (a_G+\int ds\, \rho_G)
\end{array}
\right)\,.
\ee
Let us further simplify the scenario by assuming there is no degeneracy at the scaling dimension and spins of $F_0, G_0$. Therefore $F_0$ and $G_0$ can be normalized such that $F_0=* \,G_0$, and the coefficients of the above matrix are all equal to one. From this we get a zero determinant, leading to
\be
 \left(a_F+\int ds\, \rho_F\right)\, \left(a_G+\int ds\, \rho_G\right) =1\,.
\ee
However, positivity in the infrared limit implies $a_F\, a_G \ge 1$,  eq. (\ref{ines}). 
This and the positivity of $\rho_F$ and $\rho_G$ implies 
\be
\rho_F=0\,,\quad \rho_G=0\,.
\ee
This gives correlation functions obeying the massless Klein Gordon equation, and a free theory. It is not difficult to realize that following the same calculation, the slightly more complicated case of $k=q=d/2$, where there can be mixed terms in the correlator matrix such as (\ref{priaa}-\ref{pribb}), leads again to free fields. 

Another way to say this is that as $F_0=* \,G_0$ and this is the unique non degenerate field at that dimension and spin, the fields ${\cal N}(F_0)\sim F$ and ${\cal N}(* \,G_0)\sim * \,{\cal N}(G_0)\sim * \,G$ are proportional to each other. This gives  
\be
d\, F=d\, * \, G=0 \,\,\Rightarrow\,\, \delta \, G=0\,\,\Rightarrow\,\,  \square G =(d\,\delta+ d\, \delta) \, G=0\,,
\ee
and the same is true for $F$ from $dG=0$.

We can consider the more general case in which the dimensions $d/2$ and the spins of $F,G$ at the UV are degenerate, but still have finite renormalizations. In this case we get a decomposition
\be
F=\tilde{F}+F_1\,, \qquad G=\tilde{G}+G_1\,. \label{sumFG}
\ee
The fields $\tilde{F}$ and $\tilde{G}$ are responsible for the form-symmetry at the UV,\footnote{More precisely, their UV limits $\tilde{F}_0$ and $\tilde{G}_0$ are responsible for the form-symmetry.} and can then be chosen such that $\tilde{F}=*\,\tilde{G}$. The fields $\tilde{F}, F_1, G_1$ are free and uncorrelated at the UV. If $F_1$ and $G_1$ were correlated at the UV we would have a component $F_1=* \,G_1$ that could be absorbed into the definitions of $\tilde{F}$ and $\tilde{G}$. We still have $d F=d \tilde{F}+d F_1=0$, $d G=d \tilde{G}+ d G_1=0$, but the individual fields $\tilde{F}, \tilde{G}, F_1, G_1$ could be non closed outside the fix point. It certainly results in a  highly tuned scenario. But the positivity analysis along the same lines as the calculation above gives again a free result. The details of this calculation are given in appendix \ref{fild2}.    

It rests to analyse the case when at least one of the fields have infinite renormalization. With this we mean the integral of the spectral measure is divergent, $\int \,ds\, \rho(s)=\infty$. In this case the two point function is more singular than the free one, and it is not difficult to see that $Z(\lambda)$ has to go to zero faster than the one corresponding to a free field: 
\be
\lim_{\lambda\rightarrow 0}\,Z(\lambda)\, \lambda^{-d/2}=0\,.
\ee
In particular, we will then have  
\be
\lim_{\lambda\rightarrow 0} Z_F(\lambda)\, Z_G(\lambda)\, \lambda^{-d}=0\,.
\ee
In this case the scaled fluxes 
\bea
\Phi_F^{\lambda} &=& Z_F(\lambda) \,\int_{\Sigma_F}\, F(\lambda\, x)= Z_F(\lambda) \,\lambda^{-k}\, \int_{\lambda \Sigma_F}\, F(x) \,\,\,\substack{\longrightarrow \\ \lambda\rightarrow 0}\,\,\, \Phi_{F_0}=\int_{\Sigma_F}\, F_0(x)\,,\\
\Phi_G^{\lambda} &=& Z_G(\lambda) \,\int_{\Sigma_G}\, G(\lambda\, x)= Z_G(\lambda) \,\lambda^{-q}\, \int_{\lambda \Sigma_G}\, G(x)\,\,\,\substack{\longrightarrow \\ \lambda\rightarrow 0}\,\,\, \Phi_{G_0}=\int_{\Sigma_G}\, G_0(x)\,,
\eea
are conserved and (a smeared version of them) have finite expectation values $\langle \Phi_{F_0}^2\rangle, \langle \Phi_{G_0}^2\rangle$, in  the UV limit. However,  
the commutator goes to zero at the UV since
\be
[\Phi_F^\lambda,\Phi_G^\lambda]=i \, Z_F(\lambda)\, Z_G(\lambda)\, \lambda^{-d}\rightarrow 0\,.
\ee
In other words, the UV fields $F_0, G_0$ are closed, but their mixed correlation function does not contain the linking number term. The reason is that such term does not renormalize in the QFT as shown previously, and it is erased by renormalization in the UV theory. 

If the UV field  $F_0=d \phi_0$ is exact, we have a field $\phi\rightarrow \phi_0$ in the QFT. It follows that $d \phi$ is a component of $F$. We can simply eliminate this component and redefine $F\rightarrow F-d \phi$. The new field is still closed, and still generates the same non-compact form-symmetry, because this cannot be changed by the addition of the exact field. This way we eliminate the possibility that $F_0$ is (physically) exact at the UV. The same happens for $G_0$.

On the other hand, if $F_0, G_0$ are not exact, they generate non trivial form-symmetries in the UV theory, and these form symmetries are not dual to each other.
We can define the scaling non-local unitary flux operators, in analogy with (\ref{45}-\ref{47}) as 
\be
 W^\lambda(q)=e^{i \,q\, Z_F(\lambda) \,\int d^dx\,\, \omega(x)\, F(\lambda\, x)}\,, \label{w}
\ee 
where $\delta \omega=J$ has support on a fixed topologically non contractible region $R$, 
with unit charge, see eq. (\ref{47}).  The weak limit of this non-local operator yields a non local operator {based on $R$, with charge $q$ in the UV theory
 \be
  \lim_{\lambda \rightarrow 0} W^\lambda(q)=W^0(q)=e^{i \,q \,\int d^dx\,\, \omega(x)\, F_0( x)}\,.
\ee 
 Therefore, by the general analysis described in \cite{Casini:2020rgj,Casini:2021zgr} and reviewed above, there must be dual operators $T^0(g)$ with non saturated expectation values, and commutation relations 
\be
W^0(q)\, T^0(g)= e^{i\, q\, g}\,  T^0(g)\, W^0(q)\,.\label{tio}
\ee 
These $T^0(g)$ are generated by the dual form field $* F_0$ in case  the UV is free, otherwise the operators $T^0(g)$ are of discrete charges $g\in 2\pi n$, if we  choose the $U(1)$ group to be for $q\in [0,1)$. In either scenario, out of the fix point, the $ W^\lambda(q)$ are genuine non-local operators of the theory. Then, for finite small regions, as $\lambda\rightarrow 0$, there must be a set of operators $T^\lambda(g)$ satisfying (\ref{tio}) with $W^\lambda(q)$ in the full theory too, and such that their expectation values converge to the ones of $T^0(g)$.

We could search the operators $T^\lambda(g)$ among the ones generated by fluxes of $G$, but this is impossible. If we try to keep the commutator of the two non local operators fixed as $\lambda\rightarrow 0$, in analogy with (\ref{w}) we have to take an operator of the form
\be
T^{\lambda}(g)=e^{i \,g\, Z^{-1}_F(\lambda)\, Z^{-1}_G(\lambda)\,\lambda^d\, \,\int d^dx\,\, \tilde{\omega}(x)\, Z_G(\lambda)\,G(\lambda\, x)}\sim e^{i \,g\, (Z^{-1}_F(\lambda)\, Z^{-1}_G(\lambda)\,\lambda^d)\, \,\int d^dx\,\, \tilde{\omega}(x)\, G_0(x)}\,.
\ee
This leads to a scaling of the flux of $G$ that is too little suppressed to produce an operator with non zero expectation value in the limit. 

This means the duality defects of the theory form a group larger group, $\mathbb{R}\times U(1)$ or larger, instead of just $R$. Then  $F$ is a mixing of generators of two (or more) form-symmetries. Therefore, we can reset our definition of $F$ extracting the component of larger renormalization (larger scaling dimension if there are more than one different scaling dimensions). It is clear that because there is a finite number of independent fields in the UV in a range of dimensions, the process can be continued until we get an $F$ with finite renormalization, showing there must be a free sector of the theory.

\section{Discussion}
\label{discussion}

Lagrangian formulations of neutral electrodynamics, such as non-linear electrodynamics (e.g Born-Infeld) or a photon magnetically coupled to a neutron field, and also low energy effective theories of Goldstone modes, are all non-renormalizable. In this article we have explored the question of whether such class of theories can be UV completed, or else the non-renormalizable behavior is pointing to some deeper features. To approach this problem we have first observed that this class of theories is better defined by their generalized symmetries. In particular they all share the same structure of generalized symmetries, namely non-compact form-symmetries, as properly defined in the main text. In term of these symmetries the question becomes: can a UV complete theory with non-compact form-symmetries be interacting? The analysis described in the main text shows this is not possible, and that UV completing interacting neutral electrodynamics or interacting Goldstone modes must necessarily involve breaking these symmetries. This must be due to the existence of charged operators at  a certain energy scale, that would violate the closeness of the form fields, or make them ill defined. Since the form-symmetries are well defined at the classical level, see App.~(\ref{clasical}), this obstruction can be seen as a new form of a quantum anomaly.

An important question naturally arises: at which energy scale do the electric and/or magnetic charges appear in the spectrum and destroy the generalized symmetry? If a perturbative Lagrangian formulation is at our disposal, like the one described in the introduction~(\ref{lagra}), it is natural to expect that an upper bound to the mass of these predicted charged particles is given by the appropriate dimensionful couplings appearing in the Lagrangian. But actually there is no specific reason to expect saturation of this bound, and indeed in the Standard model the associated particles appear at scales much below the bound. It would be convenient to develop a direct method for computing these masses from the infrared effective theory. These masses might be expected to show up in the structures of the correlators of the dual conserved form-fields generating the form-symmetry. We have performed preliminary explorations in perturbation theory and tentatively concluded that the symmetry is preserved in a Feynman diagramatic expansion of the effective model. Further work is required to answer this question.

The problem considered in this article was motivated and solves one of the conjectures described in \cite{Benedetti:2022zbb}. As explained in such reference, the present proven conjecture, together with another conjecture concerning the existence of local Noether currents in certain particular scenarios, strongly expands the scope of the Weinberg-Witten theorem \cite{WEINBERG198059}. In particular, if both conjectures are true, any QFT with a low energy graviton field is automatically free. This rules out potential QFT's of quantum gravity without a stress tensor. The results of this article then motivate the consideration of the conjecture concerning Noether currents described in \cite{Benedetti:2022zbb}.

On a different standpoint, our results might potentially contribute to the understanding of the completeness principle in quantum gravity \cite{Polchinski:2003bq,Banks:2010zn,Heidenreich:2020tzg,Casini:2021zgr}. This principle states that associated with any low energy gauge field on a gravity theory we have a maximum set of electric/magnetic charges consistent with the Dirac quantization condition. These set of charges completely break the generalized symmetry associated with the gauge field. A simple mechanism proving this conjecture in the context of AdS-CFT was provided in \cite{Casini:2021zgr}. In this article we have seen a partially similar phenomenon purely in a QFT scenario. Intuitively, in higher derivative theories with generalized symmetries we expect charges to appear as we climb the energy spectrum. Since gravity is essentially a higher derivative theory at the perturbative level, we might expect that arguments in the direction of the present article might help to understand the necessary appearance of charges in quantum gravity. More importantly, in consonance with the first problem mentioned above, we should be able to understand the energy scales at which these new particles appear.

\section*{Acknowledgements}
 This work was partially supported by CONICET, CNEA and Universidad Nacional de Cuyo, Argentina. The work of V. B. is supported by CONICET, Argentina. The work of H. C. is partially supported by an It From Qubit grant by the Simons foundation. The work of J.M is supported by a DOE QuantISED grantDE-SC0020360 and the Simons Foundation It From Qubit collaboration (385592).

\appendix

\section{Classical interacting theories with non-compact k-form symmetry}
\label{clasical}
In this appendix we define non-compact $k$-form symmetries in classical field theory and show no contradictions arise for interacting models. These symmetries are anomalous in the sense that these models cannot exist at the quantum level.  

Our definition of $k$-form symmetry requires the analysis of localization of observables and of commutation relations. In the classical theory observables are represented by functions in phase space and commutation relations by Poisson brackets. This forces a canonical formalism but we also need a space-time description to localize observables. This is accomplished by the covariant phase space formalism \cite{Peierls:1952cb,dewitt1967quantum,de1963dynamical}. 

The starting point is an action $S$ with local Lagrangian in terms of fields. The theory may have gauge symmetries but the action is gauge invariant. The Lagrangian  can include higher derivative terms, and anti-commuting fermion fields can also be treated \cite{de1963dynamical}.  We follow the compact notation of de Witt \cite{dewitt1967quantum} and call $\phi_i$ to the fields, where the index $i$ includes space-time coordinates. The variation of the action vanishes on the solutions of the equations of motion, and this is written
\be
\frac{\delta S}{\delta \phi_i}\,\equiv\, S_{,i} \,\dot{=}\, 0\,.
\ee
The symbol $\dot{=}$ means an equation valid on the solution of the equations of motion.

Consider a gauge invariant functional $A$ of the fields with compact support $\textrm{sup}(A)$ in spacetime. This support is the set of points such that the functional depends on the field values at those points.  This functional has two different roles. First it can be used to perturb the action. Second, evaluated on the solutions of the equations of motion the functional $A$ has the interpretation of an element of the phase space of the theory. Given an element of the phase space of the theory $A$ as a gauge invariant functional of the solutions of the equations of motion we can produce a gauge invariant functional on the fields that is unique up to terms proportional to the equations of motion themselves. This ambiguity will not be relevant in what follows.

If we perturbe the action with $S\rightarrow S+\epsilon\, A$, the change in solutions obey, to linear order in $\epsilon$, the equations  
\be
S_{,ij}\,\delta_A \phi_j \,\dot{=} -\, A_{,i}\,.\label{tyy}
\ee  
$S_{,ij}$ is a local differential operator that depends on the fields if the Lagrangian is not quadratic.  The solution of (\ref{tyy}) is not unique. There are solutions of the homogeneous equation $S_{,ij}\,\delta \phi_j \dot{=}0$ because they are infinitesimal gauge transformations around the background fields (solutions of the equations of motion), and there are also solutions of the homogeneous equation because there are physical linearized perturbations $\delta \phi_j$ that propagate in the background from past to future. One may consider special solutions obeying retarded and advanced boundary conditions   
\be
\delta_A^\pm \phi_i=0\, \hspace{.7cm} \textrm{for} \,\, x\in I^\pm(\textrm{sup}(A))\,, x\notin \textrm{sup}(A)\,,\label{lim}
\ee
where $I^\pm(X)$ is the future and past of $X$.
This does not eliminate gauge redundancy but the change in physical gauge invariant observables turns out to be also gauge invariant, defining the new phase space element  
\be
\delta_A^\pm \, B\equiv  B_{,i}\,  \delta_A^\pm \phi_i\,.
\ee
Notice that, taking aside gauge invariance, the advanced and retarded solutions are unique because the solution and all its derivatives vanish in the past or future. 

We have the reciprocity relations $\delta_A^\pm \, B=\delta_B^\mp \, A$. The Peierls/Poisson  bracket is defined as
\be
\{A,B\}=\delta_A^- \, B- \delta_A^+ \, B=\delta_A^- \, B- \delta_B^- \, A=\delta_B^+ \, A- \delta_B^- \, A\,.
\ee
This bracket obeys the usual relations, including the Jacobi property.

If two observables are space-like separated we have  $\{A,B\}=0$. This requires some qualification in the case of higher derivative Lagrangians, because the advanced and retarded perturbation moves in the background field of any solution of the equations of motion. So we have to check that this background does not enlarge the light cones for the propagation of the perturbation. This is an independent causality constraint for the classical non linear theory that we have to assume. For the Lagrangians (\ref{lagra}) and (\ref{lagra1}) this was studied in \cite{adams2006causality}. The result is that the classical theories are causal if the coefficients of the non linear terms satisfy certain positivity constraints.  

Once we have a covariant phase space, a classical form-symmetry can  be defined in analogy with the quantum case \cite{Casini:2020rgj}. Suppose we have a gauge invariant closed $k$-form $F$. As discussed in section \ref{setup}, a flux $\Phi_F=\int_{\Sigma_F} F$ constructed with $F$ commutes with all observables $O$,  having a topologically trivial support, and spatially separated from the boundary $\Gamma_F=\partial \Sigma_F$. This is because we can deform $\Sigma_F$ such as to be spatial from $O$. For $k=d-1$ we have to ask this property as an independent condition. This form-field produces a form-symmetry only when $F$ is not the exterior derivative of a gauge invariant $k-1$ form. Otherwise $\Phi_F$ can be written as an integral over $\Gamma_F$ and is not a non local operator. 

Suppose now that we have closed $k$ and $d-k$ forms $F$ and $G$. Take the phase space flux elements $\Phi_F$, $\Phi_G$, over surfaces with boundaries $\Gamma_F$ and $\Gamma_G$, linked to each other, and assume the fluxes do not commute. In this case, it is automatic that these represent form-symmetries because the forms cannot be exact in the gauge invariant space. As described in section \ref{setup}, if we move the boundaries  $\Gamma_F$ and $\Gamma_G$, keeping them linked and space-like separated, the Poisson bracket $\{A,B\}$ cannot change. This is because the change in each of the variables is an integral of observables spatially separated from the other. Therefore, we can deform the boundaries keeping them linked and keeping the commutator invariant. In particular we can move $\Gamma_F$ and $\Gamma_G$ far away together.  Then  $\{A,B\}$ has to commute with any phase space element, and has to be a number, which can be normalized to $1$. Then there is a form symmetry associated to the Abelian non-compact group $\mathbb{R}$.   
 
Suppose we have this situation for a theory where the Lagrangian has a free term that already displays the non-compact form symmetry, and an interacting term that does not ruin this dual form symmetries. To be concrete consider 
\be
L= -\frac{1}{4}\, \left(F^2+\frac{\mu}{2}\, (F^2)^2\right)\,.  
\ee 
with $F_{\mu\nu}=\partial_\mu A_\nu-\partial_\nu A_\mu$ as usual. We can change this Lagrangian to 
 \be
L= -\frac{1}{4}\, \left(F^2+\frac{\mu(x)}{2}\, (F^2)^2\right)\,,  
\ee 
where $\mu(x)$ depends on the coordinates, is constant in a region $\Lambda$ of space-time, and zero sufficiently far in space and time.  The equation of motion is 
\be
\partial_\nu \left(F^{\mu\nu} (1+\mu(x)\, F^2) \right)=0\,. 
\ee
Therefore, we have two gauge invariant closed two forms 
\be
F_{\mu\nu}\,,\hspace{.7cm} *\,(F_{\mu\nu} (1+\mu(x)\, F^2))\,.
\ee
We can apply the above reasoning to show that linked fluxes by these two forms have a constant commutator. This commutator can in particular be evaluated at spatial infinity where $\mu(x)=0$, giving us the same commutator as in Maxwell theory. The same commutator holds in the region where $\mu(x)=\mu$. We can then take the limit where $\mu(x)=\mu$ in all space-time. We see the addition of terms to the Lagrangian that do not contain charges, and consequently cannot destroy the conservation of one of the closed forms, will only deform the expression of the form, and must keep the commutator fixed. Hence, these classical models exhibit non-compact form symmetries and are interacting.  
 
Because of Groenewold's theorem \cite{groenewold1946principles} no reasonable mapping of classical phase space to  Hilbert space operators respecting the non linear structure and giving a representation of Poisson brackets is possible. Because of that, quantization requires further structures such as required by geometric quantization, deformation quantization, or the path integral. Anomalies  get in there, and in the present case we have seen there is no corresponding quantum model to these classical ones.  

\subsection{Compact case}

In this short section we look more closely into what happens when one of the symmetries is compact. Let us think in the case of a broken continuous symmetry giving place to a Goldstone boson for concreteness. The Lagrangian
\be
 L=\frac{1}{2} (\partial \psi)(\partial \psi)^*-\frac{\lambda}{4} (|\psi|^2-v^2)^2\,
\ee
 has a broken $U(1)$ symmetry. Writing $\psi=(v+\sigma) \, e^{i \phi}$ we get
\be
L= \frac{1}{2} \,(\partial \sigma)^2+\frac{1}{2} \,(\partial \phi)^2\, (v+\sigma)^2 -\frac{\lambda}{4}(\sigma^2+ 2 v\, \sigma)^2\,. \label{36}
\ee
This looks like we have a non-compact symmetry with dual closed $1$-form $\partial_\mu \phi$ and $(d-1)$-form $* \,((\partial_\mu \phi)\, (v+\sigma)^2)$, as dictated by the equations of motion. However, $\phi$ is only defined up to multiples of $2\pi$. Both currents are invariant under the shift $\phi\rightarrow \phi+ 2\pi$. In the full theory we can only smear $e^{i \int \alpha(x)\,\phi(x)}$ with $\int \alpha(x)=n$ an integer. This implies we cannot form the fluxes of $\partial_\mu \phi$ along a line with arbitrary coefficients. In the classical theory this means only quantized fluxes are allowed and it does not represent an ordinary form field. 
In the quantum theory this means the field $\phi$ or $\partial_\mu \phi$ are not Wightman fields (the same holds for $\sigma$).  In the path integral they can be represented as fields provided it is also summed over topologically non equivalent contributions for the amplitudes, where the field $\phi$ goes to $2\pi n$ at infinity rather than to zero. So the Lagrangian (\ref{36}) is not telling the theory because we have a hidden prescription for the path integral that has to be added.  However, the current that is conserved because of the equations of motion is simply
\be
((\partial_\mu \phi)\, (v+\sigma)^2)\sim i(\psi\,\partial_\mu \psi^*- \psi^* \partial_\mu \psi)\,, 
\ee 
which is a good candidate to represent a Wightman field. 

\section{Biforms and notation} 
\label{notation}
The tensor structure of the two-point function of form fields are described by biforms. These can be represented by tensors with two pairs of antisymmetric indexes. In this appendix, we will set the notation  for a  $(k|q)$  biform  $T$ with two multi-indexes $\mu,\nu$. Note that $\mu$ stands for the first set of $k$ antisymmetric indexes  $\mu_1, \mu_2, \,...\,,\,\mu_k,$ while $\nu$ describes the following $q$  antisymmetric indexes  $\nu_1, \nu_2, \,...\,,\,\nu_q$. Namely, we can write
\be 
\big[T \big]_{\mu|\nu} \equiv  \big[T \big]_{\mu_1 \mu_2 \,...\,\mu_k|\nu_1 \nu_2 \,...\,\nu_q}\,.
\ee
A nice explicit example of a biform  is the generalized metric $g^{(k)}$ introduced in (\ref{yu}). This is a $(k|k)$ biform obtained by the contraction of two antisymmetric Levi-civita tensors  as 
\be
g^{(k)}_{\mu|\nu}\equiv \frac{(-1)^{d-1}}{(d-k)!} \varepsilon_{\mu}^{\,\,\,\alpha} \varepsilon_{\nu\alpha }=\frac{(-1)^{d-1}}{(d-k)!} \varepsilon_{\mu_1 \mu_2 \,...\,\mu_k}^{\,\,\,\qquad\quad\alpha_1\cdots \alpha_{d-k}} \varepsilon_{\nu_1 \nu_2 \,...\,\nu_k\alpha_1\cdots \alpha_{d-k} }\,.
\ee 
In general, the usual differential geometry operations can act on either of the two set of indexes. By writing the operation on the left side of $T$ we denote that the operation acts on the first $k$ indexes labeled by $\mu$. For instance, the wedge product $p \wedge T$ where $p^\alpha$ is the  momentum vector yields the $(k+1|q)$ bifom given by
\be 
\big[p \wedge T \big]_{\mu | \nu} = \frac{1}{k!} \, g^{(k+1)}_{ \mu\,|\, \alpha \rho}\, p^\alpha\, T^{\rho}_{\,\,\,\nu} \equiv \frac{1}{k!} \, g^{(k+1)}_{\mu_1 \mu_2 \,...\,\mu_k\mu_{k+1}\,|\, \alpha \rho_1 \rho_2 \,...\,\rho_k}\, p^\alpha\, T^{\rho_1 \rho_2 \,...\,\rho_k}_{\,\,\,\qquad\quad\nu_1\nu_2 \,...\,\nu_q}\,.
\ee
On the other hand, if the operator is  on the right side of $T$, it acts on the second $q$ indexes labeled by $\nu$. In this case the wedge product produces a  $(k,q+1)$ biform
\be 
\big[ T  \wedge p \big]_{\mu | \nu} = \frac{1}{q!} \, g^{(q+1)}_{ \nu\,|\,  \rho \alpha}\, T^{\mu}_{\,\,\,\rho} \, p^\alpha \equiv \frac{1}{q!} \, g^{(q+1)}_{\nu_1\nu_2 \,...\,\nu_q\nu_{q+1}\,|\,  \rho_1 \rho_2 \,...\,\rho_q \alpha}\, T_{\mu_1 \mu_2 \,...\,\mu_k}^{\,\,\,\qquad\quad \rho_1 \rho_2 \,...\,\rho_q} \, p^\alpha\,.
\ee
The same ideas can be extended for other operators. The dot product acting on the left $p\cdot T $ produces a $(k-1,q)$ biform while $T\cdot p $ gives a $(k,q-1)$ biform
\be 
\big[p \cdot T \big]_{\mu |\nu} = p^\alpha \,T_{\alpha\mu|\nu} \equiv  p^\alpha \, T_{\alpha\mu_1 \mu_2 \,...\,\mu_{k-1}|\nu_1\nu_2 \,...\,\nu_q}\,.
\ee
\be 
\big[T \cdot p\big]_{\mu|  \nu} =  T_{\mu|\nu\alpha}\,p^\alpha \equiv   T_{\mu_1 \mu_2 \,...\,\mu_{k}|\nu_1\nu_2 \,...\,\nu_{q-1}\alpha}\,p^\alpha.
\ee
In addition, the Hodge dual yields a $(d-k,q)$ or a $(k,d-q)$ biform respectivelly
\bea 
\big[*T \big]_{\mu |\nu} = \frac{1}{k!} \, \varepsilon_{\mu\rho}\, T^{\rho}_{\,\,\,\nu} \equiv  \frac{1}{k!} \, \varepsilon_{\mu_1 \mu_2 \,...\,\mu_{d-k}\rho_1 \rho_2 \,...\,\rho_k}\, T^{\rho_1 \rho_2 \,...\,\rho_k}_{\,\,\,\qquad\quad\nu_1\nu_2 \,...\,\nu_q}\,,
\\
\big[T * \big]_{\mu |\nu} = \frac{1}{q!} \, T_{\mu}^{\,\,\,\rho} \,  \varepsilon_{\rho \nu}\equiv   \frac{1}{q!} \,T_{\mu_1 \mu_2 \,...\,\mu_k}^{\,\,\,\qquad\quad \rho_1 \rho_2 \,...\,\rho_q} \,  \varepsilon_{\rho_1 \rho_2 \,...\,\rho_q \nu_1 \nu_2\,...\,\nu_{d-q}}\,.
\eea
With this normalization and the metric in signature $(+, - - \,...)$ we have that
\be 
\big[**\,T \big]_{\mu | \nu} = (-1)^{(k-1)(d-1)}\,\big[T \big]_{\mu|  \nu}\,,\quad \big[T **\big]_{\mu|  \nu} = (-1)^{(q-1)(d-1)}\,\big[T \big]_{\mu | \nu}\,.
\ee
It will be  useful to define  the operator $\tilde{*}$ acting on the right side as
\be 
\big[\,T\, \tilde{*}\, \big]_{\mu | \nu} = (-1)^{q(d-q)}\,\big[T * \big]_{\mu | \nu} =    \frac{1}{q!} \, T_{\mu}^{\,\,\,\rho}\,  \varepsilon_{\nu \rho} \equiv   \frac{1}{q!} \, \varepsilon_{\nu_1 \nu_2\,...\,\nu_{d-q}\rho_1 \rho_2 \,...\,\rho_q }\,T_{\mu_1 \mu_2 \,...\,\mu_k}^{\,\,\,\qquad\quad \rho_1 \rho_2 \,...\,\rho_q}\,.
\ee
\section{A field satisfying the Klein Gordon equation is free}
\label{free}
It is well known that a field satisfying a linear equation is free. For example, the case of a scalar field satisfying the massless Klein Gordon equation $\square \phi=0$. There are several proofs of this fact in the literature. Free here means either that the commutator is a c-number or that the field is Gaussian having $n$ point functions satisfying Wick's theorem (all truncated $n$ point functions with $n>2$ vanish). Both of these properties are equivalent for a Wightman field, and characterize generalized free fields \cite{baumann1975field}. The massive case was treated first in \cite{jost,scho}. The massless case was proved in \cite{pohlmeyer1969jost}, including the subtler case of $d=2$, where the statement applies to chiral vector fields.  This last work proves the commutator has to be a c-number from support properties of the correlation functions in momentum space and the spectrum condition. The same conclusion follows from the result \cite{robinson1962support} stating that if the support of a field in momentum space does not contain a neighborhood of a spatial point then the field is a generalized free field. Indeed, the equation of motion implies the field has support on the null cone (a mass shell for the massive case) in momentum space. 
More recently, the theorem has been revisited in \cite{buchholz2022causal}, where the emphasis is in properties of the propagation of hyperbolic equations and an algebraic description.     

In this appendix we show that a simple proof follows from properties of harmonic functions in the Euclidean version of the theory. Let us first  think in a scalar field for simplicity. The Euclidean correlator $S(x,y_1,\cdots,y_n)$ is a real analytic function of $x$ except at the points $y_1,\cdots,y_n$. It is also a harmonic function in this domain because of the equations of motion $\square \phi=0$. It also falls to zero at infinity and diverges at most as a power near $y_1,\cdots,y_n$ \cite{truman1974spectrality}. By the general expression of a harmonic function in $\mathbb{R}^d-\{y_1,\cdots,y_n\}$ we have a decomposition in the different singular terms for each $y_i$ having the form \cite{axler2013harmonic}
\be  
S(x,y_1,\cdots,y_n)= \sum_{i=1}^n \sum_{m=0}^{r} \frac{q_{m,i}[x-y_i]}{|x-y_i|^{2 m+d-2}}\label{mero}\,.
\ee  
The $q_{m,i}$ are harmonic homogeneous polynomials of degree $m$ in the coordinates of $x-y_i$ whose coefficients can depend on the points $y_j\neq y_i$, and $r$ is a positive integer. This decomposition essentially follows from the uniqueness of harmonic functions once the behaviour at the boundary and at the singular points has been fixed.   

If we use this expression for $x\rightarrow  x+a, y_i\rightarrow y_i+a$, in the limit $|a|\rightarrow \infty$, it follows from the clustering property of correlators that $r=0$ and $q_{0,i}= S(y_1,\cdots,\hat{y}_i\,,\cdots, y_n)\,,$
where $\hat{y}_i$ is an omitted variable. We have normalized the two point function to be $|x-y|^{-(d-2)}$. 
This gives us the Gaussianity
\be
S(x,y_1,\cdots,y_n)=\sum_{i=1}^n \frac{ S(y_1,\cdots,\hat{y}_i\,,\cdots, y_n)}{|x-y_i|^{d-2}}\,.
\ee

For harmonic fields of arbitrary spin representation and $d>2$ Gaussianity follows in the same line.  A decomposition analogous to (\ref{mero}) holds for the massive case in terms of elementary solutions of the Euclidean Klein Gordon equation which are singular at a single point. Then, the same derivation can be also extended to fields obeying massive linear equations.

For $d=2$, using complex coordinates, it is $\square=\partial_z \partial_{\bar{z}}$. Then, for chiral operators in $d=2$ CFT's, such as currents or the stress tensor, all correlators are harmonic (holomorphic in this case) and therefore, in accordance with (\ref{mero}), the correlation functions are meromorphic. The singular structure of the OPE is enough to compute the full correlation functions in closed form. See \cite{gaberdiel2000introduction}. However, these correlators are not necessarily Gaussian and in general the two point function is not enough to determine them.  The novelty here is that the coefficients of $q_{m,i}$ can depend on $y_i$ for $d=2$, and the correlator is still harmonic in $y_i$, while this cannot happen in $d>2$. This has the effect that there are terms in $x-y_i$ in (\ref{mero}) such that $q_{m,i}$ can decay to zero in the limit $|a|\rightarrow \infty$, making this term invisible in the clustering limit.  
A chiral current is always Gaussian however.

  \section{Flux commutator from two-point functions}\label{AppFlux}
In this appendix compute the commutator between the fluxes $\Phi_F$ and $\Phi_G$ defined as integrals of the form fields $F$ and $G$ as in (\ref{fffg}). In particular, $F$ is defined to be a  $k$-form field, so it can be integrated on any $k$-dimensional oriented surface $\Sigma_F$.  For this we require a map $\varphi_F$ that describes the embedding of $\Sigma_F$ in $\mathbb{R}^d$ via 
\be 
\varphi^\mu_F ( s_1,\, s_2,\,...\,, s_k)= \Big( \varphi^1_F ( s_1,\, s_2,\,...\,, s_k)\,,\,....\,,\varphi^d_F ( s_1,\, s_2,\,...\,, s_k)\Big)\,, 
\ee
where the variables $ s_1,\, s_2,\,...\,, s_k$ parametrize the surface by taking values in a domain $S_F \subset \mathbb{R}^k$. In this context, the flux is defined as the pullback of $F$ onto $S_F$ or more explicitly 
\be 
\Phi_F (\Sigma_F) = \int_{S_F} d^k s \, F_{\mu_1 \mu_2 ...\mu_k} \big(\varphi^1_F (s_1,s_2,...,s_k), ...,\varphi^d_F (s_1,s_2,...,s_k) \big) \, \frac{\partial \varphi^{\mu_1}_F }{\partial s_1}\,\frac{\partial \varphi^{\mu_2}_F }{\partial s_2}\,...\,\frac{\partial \varphi^{\mu_k}_F }{\partial s_k} \label{fluxF}\,.
\ee

We now begin by computing the commutator of the fluxes defined over a $k$-dimensional  infinite spatial "square" $\Sigma_F^\infty$ and a $(d-k)$-dimensional one $\Sigma_G^\infty$. We chose the coordinates $(x_0,\,x_1,\,...\,,\,x_{k},\, x_{k+1}, \,...,\,x_{d-1})$ so that the surfaces are defined by
\begin{align}
\Sigma_F^\infty &\equiv \Big\{x^0=0\,, \,\,x^1 \in (-\infty, a )\,, \,\, x^2,x^3,...x^k \in \mathbb{R}\,, \,\,x^{k+1}, \,...,\,x^{d-1}=0\Big\}\,, \\
\Sigma_G^\infty &\equiv \Big\{x^0=0\,, \,\,x^1 \in (b, \infty )\,, \,\, x^2,x^3,...x^k =0\,, \,\,x^{k+1}, \,...,\,x^{d-1}\in \mathbb{R}\Big\}\,.
\end{align}
This leads to  $\Sigma_F^\infty$ and $\Sigma_G^\infty$ being parameterized  by the following maps
\be
\varphi_F^\mu = \Big(0, x^1, x^2 \,... \,,x^k, 0, 0, \,...\,,\,0\Big) \,, \quad \varphi_G^\mu = \Big(0, x^1, 0 \,... \,,0, x^{k+1}, x^{k+2}, \,...\,,\,x^{d-1}\Big) \label{par}\,.
\ee
Considering (\ref{par}) in (\ref{fluxF})  we get that  the flux of $F$ can be computed over  $\Sigma_F^\infty$ simply as
\be 
\Phi_F (\Sigma^\infty_F)= \int_{-\infty}^a dx_1 \int_{-\infty}^\infty dx_2 \,...\, \int_{-\infty}^\infty dx_k \, F_{1\, 2\,...\,k} \big(0, x^1, x^2 \,... \,,x^k, 0, 0, \,...\,,\,0\big) \,. \label{fluxF2}
\ee
To proceed  compute the expectation value of the commutator via 
\be 
\langle\big[\Phi_F (\Sigma^\infty_F),\Phi_G  (\Sigma^\infty_G)\big]\rangle = \int_{\Sigma^\infty_F}  \int_{\Sigma^\infty_G}  \langle F(x),G(y) \rangle -  \int_{\Sigma^\infty_F}  \int_{\Sigma^\infty_G}  \langle G(y),F(x) \rangle \label{excomm}\,,
\ee
where  we take into account the relevant components presented in (\ref{fluxF2}) and its analogue for $\Phi_G$. Considering that the required tensor structure takes the values 
\be 
[P^{(k)}\tilde{*}]_{1\, 2\,...\,k\,1 k+1\, ..\,d-1}=[*P^{(k)}]_{1 k+1\, ..\,d-1\,1\, 2\,...\,k} = (-1)^{k(d-k)+1} \,p_1 \,p_0
\ee
we can integrate the delta functions appearing in the integrals of the momenta $p_2,p_2,...p_{d-1}$. The resulting expression for the expectation value of the commutator is
\begin{align}
&\langle\big[\Phi_F (\Sigma^\infty_F),\Phi_G  (\Sigma^\infty_G)\big] \rangle=\frac{(-1)^{kq+1}}{\pi}\int_{-\infty}^a dx_1  \int^{\infty}_{b} dy_1 \int_0^\infty dp_0\int_{-\infty}^{\infty} dp_1  \delta(p_0^2-p_1^2)\,p_1 \,p_0 \,e^{i p_1 (x_1-y_1)}\nonumber\\
&=(-1)^{kq}\,\frac{i}{2 \pi}\int^{\infty}_{b} dy_1 \int_{-\infty}^a dx_1  \, \frac{\partial}{\partial x_1} \int_{-\infty}^{\infty} dp_1   \,e^{i p_1 (x_1-y_1)}= i\,(-1)^{kq}\, \theta(a-b)\,.
\end{align}
The dependence of the result on the Heaviside function $\theta(a-b)$ represent that the commutator is only non-vanishing when $a>b$. This means when the boundary of the  squares are linked. On the other hand, the sign  $(-1)^{kq}$ is not relevant to the result as the sign of the commutator will also change with the orientation of the surfaces. As we expect the commutator to be always a c-number, we have for linked squares extending to infinity that
\be 
\big[\Phi_F (\Sigma^\infty_F),\Phi_G  (\Sigma^\infty_G)] =\pm i\,.
\ee
This result can be  easily generalized to finite squares. The flux over a finite squares $\Sigma_F$   can be obtained by the subtracting  to the flux over   $\Sigma_F^\infty$ the flux over other infinite square  $\Sigma_F^-$ ending at $x_1<b$.  The same can be done over a dual finite square $\Sigma_G$ by subtracting the flux over $\Sigma_G^-$ ending at $x_1>a$. This means that we have 
\be\Phi_F (\Sigma_F) = \Phi_F (\Sigma^\infty_F) - \Phi_F (\Sigma^-_F)\,,\quad \Phi_G  (\Sigma_G) = \Phi_G (\Sigma^\infty_G) - \Phi_F (\Sigma^-_G)\,, \ee
where we have chosen the regions $\Sigma_F^-$  and $\Sigma_G^-$  in order to
\be [\Phi_G  (\Sigma^-_G),\Phi_F (\Sigma^-_F)] = [\Phi_G  (\Sigma^-_G),\Phi_F (\Sigma^\infty_F)]  =[\Phi_G  (\Sigma^\infty_G),\Phi_F (\Sigma^-_F)] = 0\,.
\ee
This implies that the commutator computed  over the linked finite squares $\Sigma_F$  and $\Sigma_G$  is the same as the one extending to infinity. Namely,
\be[\Phi_G  (\Sigma_G),\Phi_F (\Sigma_F)] = [\Phi_G  (\Sigma^\infty_G),\Phi_F (\Sigma^\infty_F)]  = \pm i\,. \ee
The same argument applies to other deformations of  the geometries of  $\Sigma_F$ and $\Sigma_G$ that do not  change the fact that they are linked. To sum up the commutator over linked surfaces will always be given by
\be 
\big[\Phi_F ,\Phi_G ] =\pm i\,.
\ee
It is interesting to check what happens when $k=q=d/2$, In this case, the commutator have the additional terms coming from (\ref{kkj1aa}-\ref{pribb}). We can see from (\ref{340}) that these extra terms will not change the commutator as they come from a double exterior derivative.  However, we can perform the actual computation for the case of infinite squares. The relevant components now yield
\be 
[P^{(k)}]_{1\, 2\,...\,k\,1 k+1\, ..\,d-1}=[P^{(k)}]_{1 k+1\, ..\,d-1\,1\, 2\,...\,k}= \left\{ \begin{matrix}p_1^2 & \text{if } k=1\\ p_2 p_3 & \text{if } k=2\\ 0 & \text{if } k>2\end{matrix} \right.\,.
\ee
For $k>2$ we  trivially get zero. For $k=2$ we have that replacing in (\ref{excomm}) the extra term in (\ref{hh}) writes
\begin{align}
\int_{\Sigma^\infty_F}\int_{\Sigma^\infty_G}& \int \frac{d^{4}p}{(2\pi)^{d-1}}\,\theta(p_0)\,\delta(p^2)\, P^{(2)}(p)= \\
&=\frac{1}{\pi}\int_{-\infty}^a dx_1  \int^{\infty}_{b} dy_1 \int  d^4p \,\,\, \theta(p_0)\,\delta(p^2) \,p_2\,\delta (p_2)\,p_3\,\delta (p_3)=0\,, \nonumber
\end{align}
where integrating out the delta function we get that the extra term makes no contribution to the commutator. For $k=1$ we proceed in the same way  and get
\begin{align}
\int_{\Sigma^\infty_F}\int_{\Sigma^\infty_G}& \int \frac{d^{2}p}{(2\pi)^{d-1}}\,\theta(p_0)\,\delta(p^2)\, P^{(1)}(p)= \\ \frac{1}{2\pi}&\int_{-\infty}^a dx_1  \int^{\infty}_{b} dy_1 \int d^2p\,\,\,\theta(p_0) \,\delta(p^2)\, p_1^2 \,\left(e^{i p_1 (x_1-y_1)}-e^{-i p_1 (x_1-y_1)}\right) \, =0 \,,\nonumber
\end{align}
where the zero is obtained by changing the sign of $p_1$ in the second term. To sum up we get
\be \int_{\Sigma^\infty_F}\int_{\Sigma^\infty_G} \int \frac{d^{2k}p}{(2\pi)^{d-1}}\,\theta(p_0)\,\delta(p^2)\, P^{(k)}(p)=0 
\,.
\label{dd} \ee 
As this is true for any  choice $a$ and $b$, it is valid for linked or not linked squares. Taking into account  that $dF=0$ and $dG=0$, the surfaces can be deformed once again, meaning that the result (\ref{dd}) holds for any choice of $\Sigma_F $ and $\Sigma_G$.
\section{ Renormalization group flow in AQFT }
\label{RGs}

Some of the results and derivations above have required a more careful understanding of the relation between a QFT and its UV QFT limit. During the development of this article we have explored some literature on this subject. For the benefit of the reader we have included in this appendix a review of such literature and their results. These results are expressed or use elements of the algebraic approach to QFT (the Haag-Kastler approach \cite{Haag:1963dh}), where the basic objects are algebras of bounded operators attached to space-time regions. Indeed, it is reasonable to expect that, even starting from Wightman fields, the power of the full operator content of the theory would be necessary to establish these questions. Starting from Wightman fields \cite{streater2000pct}, with minor technical changes in the axiomatic prescription, the existence of algebras affiliated with the field operators is warranted (see for example \cite{guido2011modular}). These small changes are justified from the physical point of view because the basic experimental observables are bounded operators and not point like fields. Given a theory defined in algebraic way, there is an established procedure to extract (or recover) its Wightman fields \cite{fredenhagen1981local,rehberg1986quantum,bostelmann2005phase}. For a theory generated by Wightman fields this procedure recovers the initial fields, as well as other fields that are present in the theory.           

The idea of the renormalization group can be made precise in the algebraic context through the idea of scaling algebras \cite{buchholz1995scaling}. This also defines a UV limit theory. In general this limit may not be unique, or may be classical, in the sense that all operators commute. Examples of what goes ``wrong'' can be constructed using generalized free fields (GFF). This is defined by its two point function. As happens for any field, the polynomial grow of the two point function in momentum space forces that there is a $\Delta$ such that (\ref{primo}) holds \cite{truman1974spectrality}. But (\ref{primo}) and (\ref{sec}) do not necessarily hold for the same $\Delta$. It is possible to design a Kallen-Lehmann spectral function for the correlator of the GFF such that its short distance behavior oscillates between different scaling dimensions and never actually converges.

To control the behavior of the limit theory, and eliminate these cases, it is necessary to introduce a phase space condition limiting the growth of the number of degrees of freedom at high energies. Several phase space conditions have been introduced in the algebraic context. One such phase space condition, ``uniform compactness'', ensures a unique dilatation invariant QFT limit \cite{bostelmann2010dilation}.
 Another ``microscopic phase space condition''  have been introduced in  \cite{bostelmann2005phase}, and it was proved that under this condition there exists a finite number of fields of finite spin with scaling dimensions below any fixed number $\Delta$ \cite{bostelmann2009scaling}.  
The global internal (broken or unbroken) and space-time symmetries of the QFT are kept in the scaling limit, and analogues of the renormalization functions $Z_\varphi(\lambda)$ of (\ref{eq}) can be obtained \cite{bostelmann2009scaling}. 
 The phase space condition is strong enough to allow for an operator product expansion of the fields \cite{bostelmann2005operator}.

In conclusion, a great deal of the assumptions of the present paper regarding the UV limit follow from phase space conditions that roughly speaking restricts the growth of the number of degrees of freedom in the UV. For example, it is enough that this growth is bounded above by the one of a finite number of free fields in a finite number of space-time dimensions $\ge d$. However, the matching with our requirements is not complete. For example, ref. \cite{bostelmann2009scaling} only proves that the number of independent $\varphi_0$ of the UV fix point with scaling dimension below some $\Delta$ is less than or equal to the number of linearly independent fields $\varphi$ of the QFT below the same dimension. 
\section{UV filtering of fields with finite renormalization  \label{fild2}}
In this appendix we consider the UV filtering in the case where there are more than one field in the UV with the same spin and  scaling dimension  $d/2$. Namely, if all the renormalizations are finite we can decompose the fields $F$ and $G$ as 
\be 
F=\fo+\fii \,, \quad G=\go+\gii \,.
\ee
we  can choose the fields $\fo $ and $\go$  to obey $\fo= * \,\go$  and therefore in they $IR$ their two-point functions must include a term such as (\ref{kkj1}-\ref{ppri}). 

Now we will restrict ourselfs to the study of the correlators of $\fo,\fii$. The more general expression for the two-point functions involving only $\fo$ and $\fii$ that respects the conservation of $F$ as $dF=d\fo+d\fii=0$ are given by
{\small
\begin{align}
& \langle \fo(x) \fo(0)\rangle  = \int_0^\infty ds\int  \frac{d^{d}p\,  e^{i p x}}{(2 \pi)^{d-1}}\theta(p^0)\,\delta(p^2-s)\big[( a\, \delta(s) + \rho_0(s))\ P^{(k)}\,+(-1)^k s\, \rho_1(s) g^{(k)}\big]\,, \nonumber\\
& \langle \fii(x) \fii(0)\rangle  = \int_0^\infty ds\int  \frac{d^{d}p\,  e^{i p x}}{(2 \pi)^{d-1}}\theta(p^0)\,\delta(p^2-s)\big[( b\, \delta(s) + \rho_2(s))\ P^{(k)}\,+(-1)^k s\, \rho_1(s) g^{(k)}\big]\,, \label{Fs}\\
& \langle \fo(x) \fii(0)\rangle  = \int_0^\infty ds\int  \frac{d^{d}p\,  e^{i p x}}{(2 \pi)^{d-1}}\theta(p^0)\,\delta(p^2-s)\big[( c\, \delta(s) + \rho_3(s))\ P^{(k)}\,-(-1)^{k} s\, \rho_1(s) g^{(k)}\big]\,, \nonumber
\end{align}}
where $\rho_1(s)$ appears in the three correlators such that $dF=0$, namely
\be
\langle \fo(x) dF(0)\rangle =0\,,\quad  \langle \fii(x) d F(0)\rangle=0\,.
\ee

It will be interesting to see what constraints are imposed by the postivity of the correlators. For $\langle \fo \fo \rangle$ and $\langle \fii \fii \rangle$ the positivity in the IR only reach the coefficient of the delta functions and therefore we get $a\geq 0 \text{ , and } b\geq 0$. However, at higher energies the positivity of the massive part implies that the  Kallen–Lehmann functions obey 
\be
0\leq \rho_1(s)\leq \rho_0 (s)\,,\quad 0\leq \rho_1(s)\leq \rho_2 (s)\,.
\label{pos1}
\ee
Furthermore, the positivity of the combined correlator matrix  between $\fo$ and $\fii$ yields the useful inequalities 
\be 
a\,b\geq c^2\,,\quad \rho_0(s) - 2 \rho_1(s) +\rho_2(s) \pm \sqrt{[\rho_0(s)-\rho_2(s)]^2+4[\rho_1(s)+\rho_3(s)]^2}\geq 0
\label{pos2}\,.
\ee
The last inequality has to be understood as a positivity of a matrix of measures functions.

The  fact that $\fo= * \,\go$, fixes the correlation functions involving $\go$ and $\fo$ or $\fii$. This along with $dG=0$ implies that the two-point functions are constrained to be 
{\small \begin{align}
& \langle \go(x) \go(0)\rangle = \int_0^\infty ds\int  \frac{d^{d}p\,  e^{i p x}}{(2 \pi)^{d-1}}  \theta(p^0)\,\delta(p^2-s)  \big[( a\, \delta(s) + \rho_0(s))\ P^{(q)}+ (-1)^q s\, [\rho_0(s)-\rho_1(s)] g^{(q)}\big] \nonumber \,,\\
& \langle \gii(x) \gii(0)\rangle = \int_0^\infty ds\int  \frac{d^{d}p\,  e^{i p x}}{(2 \pi)^{d-1}} \theta(p^0)\,\delta(p^2-s)   \big[( d\, \delta(s) + \rho_4(s))\ P^{(q)}+ (-1)^q s\, [\rho_0(s)-\rho_1(s)] g^{(q)}\big]  \nonumber \,,\\
& \langle \go(x) \gii(0)\rangle = \int_0^\infty ds\int  \frac{d^{d}p\,  e^{i p x}}{(2 \pi)^{d-1}}  \theta(p^0)\,\delta(p^2-s)  \big[(e\, \delta(s) + \rho_5(s))\ P^{(q)}- (-1)^q s\, [\rho_0(s)-\rho_1(s)] g^{(q)}\big]  \nonumber \,,\\
& \langle \fo(x) \go(0)\rangle = \int_0^\infty ds\int  \frac{d^{d}p\,  e^{i p x}}{(2 \pi)^{d-1}}  \theta(p^0)\,\delta(p^2-s)  \big[( a\, \delta(s) + \rho_0(s))\ P^{(k)}\tilde{*}+(-1)^{d\,k} s\, \rho_1(s)\epsilon \big] \label{manycor}\,, \\
& \langle \fii(x) \go(0)\rangle = \int_0^\infty ds\int  \frac{d^{d}p\,  e^{i p x}}{(2 \pi)^{d-1}}  \theta(p^0)\,\delta(p^2-s)  \big[( c\, \delta(s) + \rho_3(s))\ P^{(k)}\tilde{*}  -(-1)^{d\,k} s\, \rho_1(s)\epsilon \big] \nonumber \,,\\
& \langle \fo(x) \gii(0)\rangle = \int_0^\infty ds\int  \frac{d^{d}p\,  e^{i p x}}{(2 \pi)^{d-1}}  \theta(p^0)\,\delta(p^2-s)  \big[( e\, \delta(s) + \rho_5(s))\ P^{(k)}\tilde{*}-(-1)^{d\,k} s\, [\rho_0-\rho_1 + \rho_5](s) \epsilon \big] \nonumber\,, \\
& \langle \fii(x) \gii(0)\rangle = \int_0^\infty ds\int  \frac{d^{d}p\,  e^{i p x}}{(2 \pi)^{d-1}}  \theta(p^0)\,\delta(p^2-s)  \big[( f\, \delta(s) - [\rho_0+\rho_3+\rho_5])\ P^{(k)}\tilde{*}+(-1)^{d\,k} s\, [\rho_0-\rho_1 + \rho_5] \epsilon \big]\,.  \nonumber 
\end{align}}
Note that the positivity in the IR limit implies that $f\geq 0\,,\,\,ad\geq e^2\,,\,\, \text{and }\, bd\geq f^2$. Moreover,  the positivity of single correlators $\langle \go\go\rangle$ and $\langle \gii\gii\rangle$ is analogous to (\ref{pos1}). The  mixed correlator matrix of $\go$ and $\gii$, in analogy with (\ref{pos2}), gives the new constraint
\be 
 -\rho_0(s) + 2 \rho_1(s) +\rho_4(s) \pm \sqrt{[\rho_0(s)-\rho_4(s)]^2+4[\rho_0(s)-\rho_1(s)+\rho_5(s)]^2}\geq 0\,.
\label{pos3}
\ee

The straight forward computation of the two point function $\langle F G\rangle$,  using (\ref{manycor}), picks up only a massless contribution. This is
\begin{align}
\langle F(x) G(0)\rangle &=  \langle \fo(x) \go(0)\rangle+ \langle \fii(x) \go(0)\rangle + \langle \fo(x) \gii(0)\rangle+ \langle \fii(x) \gii(0)\rangle\nonumber \\
&=(a+c+e+f)\int  \frac{d^{d}p}{(2 \pi)^{d-1}}\, e^{i p x}\,  \theta(p^0)\,\delta(p^2-s) \, P^{(k)}\tilde{*}\,(p)\label{FG}\,.
\end{align}
Since this correlator does not renormalize, in the UV $\langle F G\rangle_{UV}$ is equivalent to $\langle\fo\go \rangle_{IR}$, which implies that  
\be 
c+e+f=0 \label{uv1}\,.
\ee
In addition, we ask that the fields $\fii$ and $\gii$ are uncorrelated to $\fo$ in the $UV$.  Meaning that we have  $\langle \fo \fii \rangle_{UV},\langle \fo \gii\rangle_{UV}=0$, or 
\be 
 c+\int_0^\infty \rho_3(s)\, ds =0\, ,\quad e+\int_0^\infty \rho_5(s)\, ds =0\,. \label{uv2}
\ee
The last $UV$ requirement is that $\fii$ and $\gii$ are not correlated in such a limit.  This implies no loss of generality, as if they  were correlated we would have a component $F_1=*\,G1$ that could be absorbed into the definitions of $\fo$ and $\go$. The fact that $\langle \fii\gii\rangle_{UV}=0$ gives
\be 
 f-\int_0^\infty (\rho_0(s)+\rho_3(s)+\rho_5(s))\, ds =0\ \label{uv3}\,.
\ee
Replacing (\ref{uv1}) and (\ref{uv2}) in (\ref{uv3}) we get $\int ds\, \rho_0(s) = 0$. In light of (\ref{pos1}), this means that $\rho_0(s) = 0$ and, also, $\rho_1(s) = 0$. Then, the constraints (\ref{pos2}) and (\ref{pos3}) imply that $\rho_{3}(s),\rho_{5}(s)=0$, as well as $c,e,f=0$. The final form of the non-vanishing correlators outside the UV is
\begin{align}
&\langle \fo(x) \fo(0)\rangle  =  \int  \frac{d^{d}p }{(2 \pi)^{d-1}} \,e^{i p x}\,\theta(p^0)\,\delta(p^2)\, P^{(k)}\label{F0F02}\\
&\langle \go(x) \go(0)\rangle  = \int  \frac{d^{d}p }{(2 \pi)^{d-1}} \,e^{i p x}\,\theta(p^0)\,\delta(p^2)\, P^{(q)}\label{G0G02}\\
&\langle \fo(x) \go(0)\rangle  = \int  \frac{d^{d}p }{(2 \pi)^{d-1}} \,e^{i p x}\,\theta(p^0)\,\delta(p^2)\, P^{(k)}\tilde{*}\label{F0G02}\\
&\langle \fii(x) \fii(0)\rangle  = \int_0^\infty ds\int  \frac{d^{d}p}{(2 \pi)^{d-1}}\,  e^{i p x}\,\theta(p^0)\,\delta(p^2-s)\,( b\, \delta(s) + \rho_2(s))\,\ P^{(k)} \label{F1F12}\\
&\langle \gii(x) \gii(0)\rangle  = \int_0^\infty ds\int  \frac{d^{d}p}{(2 \pi)^{d-1}}\,  e^{i p x}\,\theta(p^0)\,\delta(p^2-s)\,( d\, \delta(s) + \rho_4(s))\,\ P^{(q)} \label{G1G12}
\end{align}
where we have also fixed $a=1$. This leads to $\fo$ and $\go$ being free massless fields that are related with the form symmetry by (\ref{F0G02}). Also, the Hilbert space generated by these fields is in tensor product to the one generated by $\fii$ and $\gii$.

\bibliographystyle{utphys}
\bibliography{EE}

\end{document}